\documentclass[prd,reprint,superscriptaddress,floatfix,nofootinbib,nobibnotes,amsmath,amssymb,aps]{revtex4-2}

\usepackage{aas_macros}
\usepackage{graphicx} 
\usepackage{amssymb}
\usepackage{amsmath}
\usepackage{subfigure}
\usepackage{enumitem}
\usepackage{color}
\usepackage{siunitx}
\usepackage{parskip}
\newcommand{\msun}{\,${\rm M}_\odot$\,}

\begin{document}
\title{Gravothermal collapse of Self-Interacting Dark Matter halos \\as the Origin of Intermediate Mass Black Holes in Milky Way satellites}

\author{Tamar Meshveliani}%
\email[e-mail:]{tam15@hi.is}
\affiliation{Centre for Astrophysics and Cosmology, Science Institute,
University of Iceland, Dunhagi 5, 107 Reykjavik, Iceland}
\author{Jes\'us Zavala}
\affiliation{Centre for Astrophysics and Cosmology, Science Institute, University of Iceland, Dunhagi 5, 107 Reykjavik, Iceland}
\author{Mark R. Lovell}
\affiliation{Centre for Astrophysics and Cosmology, Science Institute, University of Iceland, Dunhagi 5, 107 Reykjavik, Iceland}

\date{\today}

\begin{abstract}
Milky Way (MW) satellites exhibit a diverse range of internal kinematics, reflecting in turn a diverse set of subhalo density profiles. These profiles include large cores and dense cusps, which any successful dark matter model must explain simultaneously.
A plausible driver of such diversity is self-interactions between dark matter particles (SIDM) if the cross section passes the threshold for the gravothermal collapse phase at the characteristic velocities of the MW satellites. In this case, some of the satellites are expected to be hosted by subhalos that are still in the classical SIDM core phase, while those in the collapse phase would have cuspy inner profiles, with a SIDM-driven intermediate mass black hole (IMBH) in the centre as a consequence of the runaway collapse. We develop an analytical framework that takes into account the cosmological assembly of halos and is calibrated to previous simulations; we then predict the timescales and mass scales ($M_{\rm BH}$) for the formation of IMBHs in velocity-dependent SIDM (vdSIDM) models as a function of the present-day halo mass, $M_0$. Finally, we estimate the region in the parameter space of the effective cross section and $M_0$ for a subclass of vdSIDM models that result in a diverse MW satellite population, as well as their corresponding fraction of SIDM-collapsed halos and those halos' inferred IMBH masses. We predict the latter to be in the range $0.1-1000$~M$_\odot$ with a $M_{\rm BH}-M_0$ relation that has a similar slope, but lower normalization, than the extrapolated empirical relation of super-massive black holes found in massive galaxies.

\end{abstract}


\maketitle

\section{Introduction}
\label{sec_intro}

The cold dark matter (CDM) model is highly successful at explaining observations of the large-scale structure of the Universe (e.g. \cite{Springel2005}). However, it has challenges in matching observations on small scales, such as in the regime of dwarf galaxies (for a recent review see e.g \cite{BBK2017}). Observationally, these challenges have been established prominently for dwarf galaxies in the Local Group, and particularly within the Milky Way (MW) satellites. For instance, the dynamical mass is dominated by dark matter (DM) in the inner region of several bright MW satellites, yet is low compared to the inner densities of the plausible subhalo hosts of MW-analogues found in collisionless CDM simulations; this is the {\it classical} too-big-to-fail (TBTF) problem \cite{Boylan-Kolchin2011, Boylan-Kolchin2012}. Another recurrent challenge is that several of the MW satellites are best explained by density profiles of constant density, known as ``cored'' density profiles, rather than the steep inner density slope found in CDM simulations, referred to as ``cuspy'' profiles; satellites that are reported to have cored profiles include Fornax, Sculptor, Crater II and Antlia II \cite{WP2011,Agnello2012,Breddels2013,Torrealba1, Torrealba2}. 
Overall, it is now well established that the MW satellites have a diverse range of internal kinematics, which is likely associated with a subhalo population that exhibits a considerable diversity of inner density profiles, from large cores to dense cuspy systems \cite{Fattahi2018,Errani,Read2019,Zavala2019,Kaplinghat2019}. This diversity is analogous to that of rotation curves observed in higher mass, gas-rich dwarf galaxies \cite{Oman2015,Santos2020}. 

It is important to emphasize that such dwarf-scale challenges are only insurmountable within CDM if no other physical mechanisms related to gas/stellar ({\it baryonic}) physics are considered. There are in fact several baryonic processes that are known to exist that can alleviate these challenges. 
For instance, supernova feedback can inject energy into the inner DM halo, reducing its density \cite{Navarro96cores}. If impulsive enough, this is an efficient and irreversible cusp-core transformation mechanism in dwarf galaxies \cite{PG2012,Burger2019,Burger2021}. In addition, tidal forces on the satellite by the MW DM halo and the MW disk can effectively lower the densities of MW subhalos if their orbits pass sufficiently close to the disk \cite{Zolotov2012}. The diverse orbits of the MW satellites combined with this effect enhance the diversity of inner DM densities relative to the CDM-only expectations \cite{GK2019}. However, how efficient these processes are in creating the observed diversity of the MW satellite population remains uncertain since, for example, the impact of supernova feedback is expected to be small in very faint DM-dominated systems with low stellar-mass ratios \cite{DC2014}. On the other hand, it has been argued that the tidal field of the MW system might not be strong enough to explain the extremely low densities of bright satellites such as Crater II and Antlia II \cite{Errani2022,Lovell2022}.

One exciting possibility is that the properties of the MW satellites provide clues about the DM nature beyond CDM. In particular, if DM particles have strong self-interactions, 
they can impact the non-linear evolution of halos, significantly reducing their inner densities \cite{Spergel}. Modified N-body simulations that incorporate self-interacting DM (SIDM) have shown that the collisions experienced by DM particles with each other lead to significant momentum exchange. This process effectively transfers heat from the dynamically hot outer regions of the halo to the colder central regions, thus lowering the central density of halos and creating constant (isothermal) density cores \cite{Dave, Colin, Vogelsberger, Rocha, Dooley, Robles,Tulin2018, Vogelsberger2019}. SIDM models can create sizeable DM cores and alleviate the {\it classical} TBTF problem if the transfer cross section per unit mass, $\sigma_{\rm T}/m_\chi$, is $\gtrsim 1$~$ \rm cm^2 g^{-1}$ at the characteristic scales/velocities of MW satellites $\lesssim \rm 50 km/s$ \cite{Zavala2013}. It is also possible to alleviate significantly the diversity of rotation curves in higher mass dwarf galaxies (characteristic velocities $>50\rm km/s$) if $\sigma_{\rm T}/m_\chi\gtrsim 2-3$~$ \rm cm^2 g^{-1}$ \cite{Creasey2017,Ren2019}. 

Constant cross section SIDM models have been constrained more strongly at large scales/velocities. Particularly, $\sigma_{\rm T}/m_\chi$ is required to be $\lesssim0.1-1$~cm$^2$g$^{-1}$ at the scale of clusters based on gravitational lensing, X-ray morphology, and dynamical analysis in cluster mergers \cite{Robertson2017,Robertson2019,Harvey2019,Andrade2022,Eckert2022,Shen2022}. At scales corresponding to massive elliptical galaxies, previous constraints based on X-ray morphology have been shown to be weaker than anticipated by DM-only simulations \cite{Peter2013} once baryonic effects are included. Current simulations including baryons have shown that $\sigma_{\rm T}m_\chi\sim1$~cm$^2$g$^{-1}$ is consistent with the morphologies of elliptical galaxies \cite{Despali2022}. With such constraints at larger scales, a constant cross section SIDM model is already only narrowly viable as an alternative to CDM to explain the properties of dwarf galaxies. Recent developments regarding the diversity in the inner densities of the MW satellites virtually rule out this possibility since with such a low cross section, it is not possible to generate very high density satellites such as Willman~I \cite{Valli2018,Read2018,Zavala2019,Kim2021}.

Remarkably, what is needed for SIDM models to remain an interesting, viable alternative to CDM is to have even larger cross sections ($>10$~cm$^2$g$^{-1}$) at the scale of the MW satellites in order to trigger the gravothermal collapse phase (see below). Such large cross sections can be naturally accommodated by particle models with a velocity-dependent cross section (e.g. through Yukawa-like interactions, see e.g. \cite{Feng2009,Buckely2010,Loeb2011}), where DM behaves as a collisional fluid on small scales and is essentially collisionless at cluster scales. 
Long after the core-formation phase, further DM particle collisions lead to heat outflow from the hotter inner region to the colder outskirts of the halo. Since gravitaionally bound systems have negative specific heat, mass/energy is continuously lost from the inner region, while the density and temperature continue to grow in a runaway instability that drives the collapse of the inner core. This phenomenon is known as the gravothermal catastrophe \cite{Lynden-Bell} and is observed in globular clusters, where the collapse is mainly halted by the formation of binary stars, which act as energy sinks \cite{Hut1992}. For SIDM halos, the physical mechanism is the same, but without the formation of bound DM states to act as energy sinks, the collapse continues, eventually reaching a relativistic instability that results in the formation of a black hole
\cite{Colin, Balberg01, Balberg02, Koda2011,Pollack}. 
If the core-collapse phase has been reached at the scales of the MW satellites, then the SIDM predictions become radically different with some of the satellites expected to be hosted by (sub)halos with SIDM cores, while those in the collapse phase would have cuspy (collapsed) inner DM regions \cite{Zavala2019}.

Given the problems with constant cross section SIDM models mentioned above, it has been argued recently that such models could be reconciled with the MW satellite population by suggesting that the collapse phase might be accelerated in the host (sub) halos of MW satellites by mass-loss via tidal stripping \cite{Nishikawa}, since mass-loss enhances the negative temperature gradient in the outskirts of the (sub)halo and makes the heat outflow more efficient. Accelerated core-collapse has been invoked to explain the diversity of the MW’s dwarf spheroidal galaxies in constant cross section models with $\sigma_{\rm T}/m_\chi\gtrsim 2-3$~$ \rm cm^2 g^{-1}$ \cite{Kahlhoefer2019,Nishikawa,Kaplinghat2019,Sameie2020}. However, Ref.~\cite{Zeng2022} recently simulated SIDM subhalo satellites as they orbit the MW system and found that energy gain due to collisions between particles in the subhalo and the host instead inhibits core-collapse in subhalos.

Another study, Ref.~\cite{Silverman}, showed that subhalos in models with constant cross sections between 1 and 5~$\rm cm^2g^{-1}$ are not dense enough to match the densest ultra-faint and classical dwarf spheroidal galaxies in the MW, and 5~$\rm cm^2g^{-1}$ is not sufficient to enforce collapse even with the tidal effect of a MW disk and bulge. This seemingly closes the last possibility for velocity-independent SIDM models (see also discussion in Section~\ref{sec_tidal-stripping} below). On the other hand, this result motivates the exploration of velocity-dependent SIDM models, where recent full cosmological simulations with a specific benchmark model \cite{Zavala2019,Turner} have shown that cross sections $\gtrsim 50$~$ \rm cm^2/g$ at velocities $\lesssim 30$~$ \rm km/s$ naturally result in a diverse bimodal population of MW satellites, predicting both cuspy, high velocity dispersion subhalos, consistent with dense systems (particularly ultra-faint satellites), and cored, low velocity dispersion subhalos, consistent with brighter low-density satellites. These results have been confirmed and expanded to generic velocity-dependent SIDM models by the recent cosmological simulation suite TangoSIDM \cite{Correa2022}.

In this work, we adopt the benchmark SIDM model presented in \cite{Zavala2019,Turner} to explore the consequences of gravothermal collapse for the formation of intermediate mass black holes (IMBHs) in the MW satellite population. Our goal is twofold: (i) to compile a simple analytical framework (calibrated to the simulations in \cite{Zavala2019,Turner}) that provides predictions for the formation timescales and mass scales of IMBHs in SIDM halos under arbitrary velocity-dependent cross sections, and (ii) to provide the range of IMBH masses that is expected given the plausible range of cross sections that produce a diverse MW satellite population, i.e., a bimodal -- core-cusp -- satellite distribution.

This paper is organised as follows. In Section~\ref{sec_SIDM-halos}, we describe our model for the evolution of SIDM halos. We start with our adopted primordial halo density profile and the concentration--mass relation, describe our computation of the threshold time for the cusp-core transformation, and finally 
estimate the timescales and masses of IMBHs expected in the SIDM model due to gravothermal collapse. We also include the impact of tidal stripping. In Section~\ref{sec_results}, we present our results, discuss how they are impacted by the various properties of the model, and put our work in the context of other related studies. Finally, we draw conclusions in Section~\ref{sec_conclusions}.
\section{Gravothermal collapse in SIDM halos}\label{sec_SIDM-halos}
Our goal in this section is to follow the relevant stages in the evolution of an SIDM halo: i) formation of the progenitor cuspy (i.e. CDM-like) halo, ii) development of the central core and iii) gravothermal collapse of the core and formation of the black hole. 
In addition, we discuss how tidal stripping might affect the gravothermal collapse timescale. 
\subsection{Cosmic evolution of SIDM halos}\label{cosmic}

In an SIDM halo where thermalization occurs due to close, rare interactions with large momentum transfer, a relaxation time can be defined due to self-scattering at the characteristic radius\footnote{From here on in, we assign the characteristic radius to the scale radius of the halo, which for the NFW profile is equal to $r_{-2}$, the radius at which the logarithm slope of the profile is $-2$; see Section~\ref{sec_NFW}.} $r_{-2}$, which is given by:
\begin{equation}\label{t_relax}
    t_{\rm r} = \dfrac{\lambda}{a\sigma_{\rm vel}},
\end{equation}
where $\sigma_{\rm vel}$ is the characteristic velocity dispersion, $a=\sqrt{16/\pi}$ for hard-sphere scattering of particles with a Maxwell–Boltzmann velocity distribution \cite{Balberg01} and $\lambda^{-1} = \rho(r_{-2}) \sigma_{\rm T}/m_\chi$ is the mean free path, which is inversely proportional to the local density $\rho(r_{-2})$ and the cross section per unit mass $\sigma_{\rm T}/m_\chi$ (evaluated at the characteristic velocity $\sigma_{\rm vel}$ in the case of velocity-dependent SIDM models). Therefore, the scattering rate (mean free path) is higher (shorter) in denser regions. Within the region where the age of the inner halo is comparable to the relaxation time, self-scattering has a significant impact on the inner DM structure turning the cusp into a core. 

In CDM, where DM is collisionless, the velocity dispersion peaks near the scale radius, $r_{\rm -2}$. By contrast, in SIDM elastic scattering leads to momentum exchange between DM particles,
which, given the 
positive gradient of the velocity dispersion profile within $r_{\rm -2}$, 
effectively results in heat 
transfer from the outside-in, up to the radius where the velocity dispersion peaks. As a result, a central isothermal core is formed, 
which continues to grow until it is roughly the size of the scale radius and thus reaches a quasi-equilibrium state.
After core formation, subsequent collisions lead to 
momentum/energy flow from the center to the outskirts of the halo, where the velocity dispersion profile has a negative slope. Heat loss in the core results in the infall of DM particles to more tightly bound orbits, where they experience more interactions and are heated further due to the negative heat capacity of the self-gravitating system; a similar phenomenon occurs in globular clusters, \cite{LB1968}. Without energy sinks, the core suffers a runaway instability, transforming the core into an ever denser cusp, which ultimately results in the formation of a black hole \cite{Balberg01}.

An SIDM halo undergoes gravothermal collapse in a timescale $t_{\rm coll}\approx 382t_{\rm r}$, as described in Section~\ref{sec_BH-formation}. The relaxation time depends on the halo mass and time of assembly/formation (described in Sections~\ref{sec_NFW}$-$ and \ref{sec_redshift}) as well as the SIDM cross section at the characteristic velocity of the halo (Section~\ref{sec_cross section}).

\subsection{Primordial density profile} \label{sec_NFW}

We assume that in the SIDM cosmology DM assembles into spherical self-gravitating halos in virial equilibrium, with a {\it primordial} structure that is the same as that of CDM halos. This is a reasonable assumption at sufficiently high redshift when the average number of collisions in the center of halos is still well below one per Hubble time, and thus the structure of the halo has been affected only minimally. Cosmological simulations have shown that DM core sizes are only a small fraction of their value at $z=0$ when the Universe is around 1 Gyr old ($z\sim5$), e.g. \cite{Vogelsberger2014}.

The spherically averaged density profiles of equilibrium collisionless CDM halos are well approximated by a two-parameter formula known as the Navarro, Frenk \& White (NFW) profile \cite{Navarro96, Navarro97}:
\begin{equation}
    \rho_{\rm NFW}(r) = \rho_{\rm crit}\dfrac{\delta_{\rm char}}{r/r_{-2}(1+r/r_{-2})^2}, 
    \label{eq:NFW}
\end{equation}
where $r_{-2}$ is the radius at which the logarithmic slope of the profile is $-2$, $\rho_{\rm crit}$ is the critical density of the Universe and the characteristic overdensity $\delta_{\rm char}$ is given by:
\begin{equation}
    \delta_{\rm char} = \dfrac{200}{3}\dfrac{c^3}{k(c)},
\end{equation}
where $k(c) =\ln(1+c)-c/(1+c)$ and the concentration $c$ is defined as $c=r_{200}/r_{-2}$ with $r_{200}$ being the virial radius, which is defined in this work as the radius where the mean density of the halo is 200 times $\rho_{\rm crit}$.

\subsection{Concentration-Mass relation and formation redshift} \label{sec_CM-relation}

The NFW profile is to first order a one free parameter profile since the virial mass of the halo and its concentration are strongly correlated, with a $1\sigma$ scatter in $\log{c}$ of order 0.1 \cite{Navarro97,Bullock2001}.
We use the concentration-mass relation modeled in \cite{Ludlow2014,Ludlow2016}, where the authors link the enclosed mass profile of a halo at a given time with the prior mass aggregation history of the halo. In particular, following \cite{Ludlow2016}, we can define an {\it assembly/formation} redshift of a halo of mass $M_0$ at a redshift $z_0$ as the redshift $z_{-2}$ when the enclosed mass within $r_{-2}$ at $z_0$, $M_{-2}$, was first assembled into progenitors more massive than a certain fraction $f$ of $M_0$.
$M_0$ is defined as the mass within the virial radius $M_0 = (4\pi/3)r_{200}^3 200\rho_{\rm crit}$. The virial mass of the halo at $z_{-2}$ is equal to $M_{-2}$ and can be computed from the assembly history:
\begin{equation}
M_{-2} = M_0 \times
    \operatorname{erfc} \left(\dfrac{\delta_c(z_{-2})-\delta_c(z_0)}{\sqrt{2(\sigma^2(f\times M) - \sigma^2(M))}}\right) 
    \label{eq:M2},
\end{equation}
The expression in parentheses on the right hand side corresponds to the collapsed mass fraction in Extended Press-Schechter theory \citep{Lacey1993},
where $\delta_{\rm c}(z_{-2})= \delta_{\rm c}/D(z)$ is the redshift dependent critical density for collapse with the linear growth factor $D(z)$, and
$\sigma(M)$ being the rms mass variance. For the NFW profile, the mass is connected to the concentration by:
\begin{equation}\label{M_2}
     \dfrac{M_{-2}}{M_0} =  \dfrac{k(1)}{k(c)},
\end{equation}
\begin{equation}\label{rho_2}
     \dfrac{\left<\rho_{-2}\right>}{\rho_{\rm crit}(z_{-2})} =
     200 c^3 \dfrac{k(1)}{k(c)}.
\end{equation}
The key assumption in the model is that the mean density inside $r_{-2}$ 
is directly proportional to the critical density of the Universe at an assembly redshift $z_{-2}$:
\begin{equation}
    \dfrac{\left<\rho_{-2}\right>}{\rho_{\rm crit}(z_{-2})} = C \left(\dfrac{H(z_{-2})}{H(z_0)}\right)^2,
    \label{eq:mean-density}
\end{equation}
 where $C$ is a free parameter. Throughout this paper we use $f=0.02$ and $C=575$ \cite{Bohr}. Inserting Eqs.~\ref{M_2}$-$\ref{rho_2} into Eq.~\ref{eq:M2}, we have a transcendental equation for the formation redshift $z_{\rm form}=z_{-2}$ as a function of $M_0$, which can then be used to obtain the concentration $c$.
\subsection{Threshold time for the cusp-core transformation}\label{sec_redshift}
As a benchmark case, we set the halo formation time $z_{-2}$ of an SIDM halo extant at the present day to be the threshold epoch at which the cusp-core transformation begins, $z_{\rm cc}=z_{\rm form}=z_{-2}$. At this epoch, we assume that the SIDM halo has an NFW profile with a virial mass equal to the enclosed mass within 
$r_{-2}$ at $z_0=0$, $M(z_{-2})=M_{-2}\vert_{z_0}$. 
The concentration of this {\it primordial} SIDM halo is calculated by repeating the method described in Section~\ref{sec_CM-relation}, but this time setting $z_0=z_{-2}$.
The range of $z_{-2}$ values for the range of present-day halo masses that we are interested in, $10^8\leq M_0\leq10^{12}$\msun, is given by $6.1\geq z_{-2}\geq 3$. As we noticed earlier, given this relatively high redshift range, our choice of setting $z_{\rm cc}=z_{-2}$ is reasonable because the effect of collisions in the inner halo is minimal at early times.

The next step is to develop a method to calculate the relevant timescale for gravothermal collapse (Section \ref{sec_BH-formation}), for which we build a simplified model in which the evolutionary stages of the SIDM halo occur in isolation. This approach is somewhat different to the full cosmological setting, where halo mergers are an active mechanism of halo growth with transitory stages that affect the inner centre of the halo. Although the cuspy NFW profile of CDM halos is resilient to merger activity \cite[e.g.][]{K2006}, the situation might in principle be more complex in an SIDM scenario with gravothermal collapse for the following reason. In the standard SIDM model without core-collapse, the merger between a small halo with a larger one is that of two shallow (core-like) profiles with the smaller one having progressed further in its core development since it forms earlier; the result of this merger is a DM profile that is also cored \cite{BK2004}. Thus, we would naively expect that halo mergers will not delay the cusp-core transformation. However, cosmological mass infall in general might delay the core-collapse phase by pumping energy into the central region to stabilize the core \cite{Ahn2005}. Moreover, in a velocity-dependent SIDM halo with a sharp difference between the cross section of low-mass halos to that of large mass halos, the former are expected to go through the cusp-core-collapse stages much faster than the latter, resulting in a scenario in which mergers between low-mass core-collapsed (cuspy) halos and high-mass cored halos are possible. This has the potential to delay the core-collapse phase. 

Since our goal is to provide a simple, first-order estimate for the black hole formation time, rather than a comprehensive calculation, we assume that in a relatively extreme scenario, a significant merger would reset the clock for the cusp-core-collapse stage. For this event we adopt the last major merger (LMM), which we define as a mass ratio of 10:1 or higher between the two merging halos, and we label the corresponding redshift as $z_{\rm LMM}$.
In order to calculate $z_{\rm LMM}$, we use the fitting formula for the mean merger rate $dN_{\rm m} /d\xi /dz$ -- in units of
mergers per halo per unit redshift per unit of mass ratio $\xi$ -- for a halo of mass $M(z)$ at redshift $z$ obtained from the combined Millennium and Millennium~II data sets in \cite{Fakhouri}: 
\begin{equation}
\begin{split}
	\dfrac{dN_{\rm m}}{d\xi dz}(M(z), &\xi, z)= \\
	&A\left(\dfrac{M(z)}{10^{12} \rm{M_\odot}}\right)^\alpha \xi^\beta \exp\left[\left(\dfrac{\xi}{\tilde{\xi}}\right)^\gamma\right] \times (1+z)^\eta,
\label{eq:fit} 
\end{split}
\end{equation}
where the best-fit parameters are $(\alpha,\beta,\gamma,\eta)=(0.133, -1.995, 0.263, 0.0993)$ and $(A,\tilde\xi)=(0.0104, 9.72\times10^{-3})$.
The mass $M(z)$ is given by integrating the mean mass growth rate of halos, taken from \cite{Fakhouri}:
\begin{equation}
\begin{split}
\dfrac{dM}{dt}=\left<\dot{M}\right>_{\rm mean} = ~&46.1 \rm{M_\odot} {\rm yr}^{-1} \left( \dfrac{M}{10^{12} \rm{M_\odot}} \right)^{1.1} \\
& \times (1 + 1.11 z) \sqrt{\Omega_{\rm m} (1+z)^3 + \Omega_\Lambda}.
\end{split}
\label{eq:M_mean}
\end{equation}
where $\Omega_m$ and $\Omega_\Lambda$ are, respectively, the DM and dark energy density parameters evaluated at the present day.

The cumulative number of mergers $N_{\rm m}(\xi_{\rm min},M_0,z_0,z)$ for a halo of mass $M_0$ at redshift $z_0$ is then given by:
\begin{equation}
    N_{\rm m}(\xi_{\rm{min}}, M_0, z_0, z)= \int_{z_0}^{z}dz \int_{\xi_{\rm{min}}}^{1}d\xi\dfrac{dN_{\rm m}}{d\xi dz}[M(z), \xi, z].
\label{eq:merger}
\end{equation}
where we use the minimum mass ratio for a major merger to be $\xi_{\rm min} = 0.1$. 
When the above equation equals to 1, meaning that the halo experienced one major merger event, we find the corresponding $z_{\rm LMM}$ for given halo of mass $M_0$; we therefore only consider the properties of halos extant at $z_0=0$. 

\begin{figure}
    \includegraphics[width=\linewidth, clip=True]{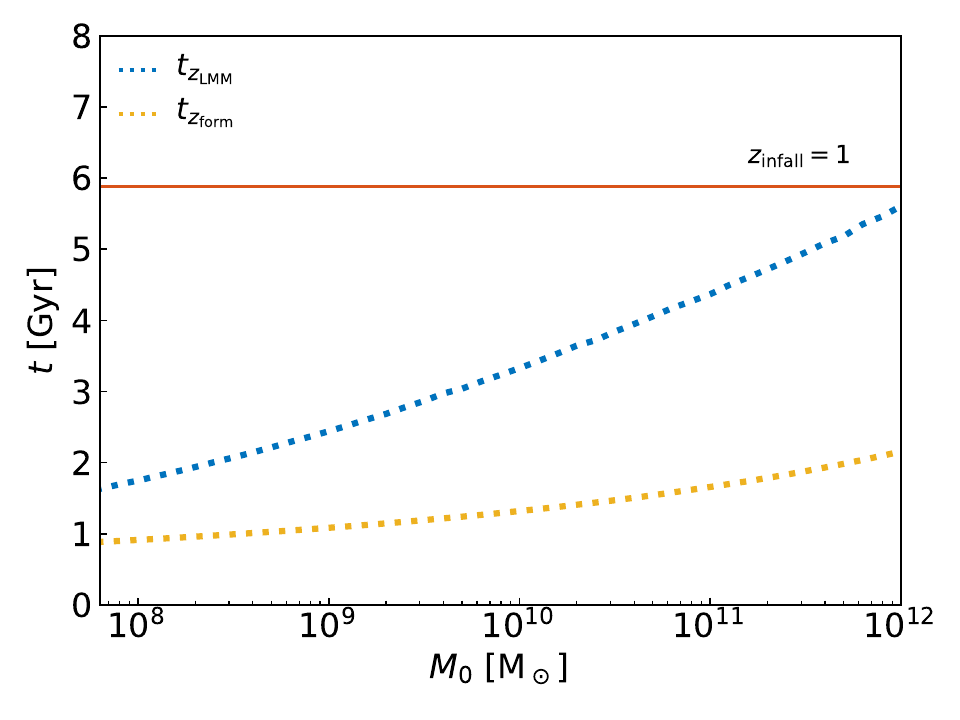}
\caption{Cosmic time for the cusp-core transformation threshold of SIDM halos as a function of the present-day halo mass $M_0$. The blue dotted line is the function of $z_{\rm LMM}$ and the yellow dotted line is the function of $z_{\rm form}$. The horizontal line indicates the {\it infall} time of subhaloes, which we fix to $z_{\rm infall}=1$ (see Section~\ref{sec_tidal-stripping}).} 
\label{fig:cosmic_time}
\end{figure}

Having adopted all these considerations, we assume that a viable range for the threshold epoch of the cusp-core transformation is given by $z_{-2}<z_{\rm cc}<z_{\rm LMM}$. The corresponding cosmic time for this epoch is given by:
\begin{equation}
t(z)=t_{0}\int_{0}^{1/(1+z)}\dfrac{da}{\dot{a}} = \dfrac{2\sinh^{-1}\left({\sqrt{\dfrac{\Omega_{\Lambda}}{\Omega_{\rm m}}} (1+z)^{-3/2}}\right)}{3H_0\sqrt{\Omega_{\Lambda}}}.
\label{eq:cosmic_time} 
\end{equation}
where $t_0$ is the age of the Universe and $H_0$ is the Hubble parameter. 
For reference, Fig.~\ref{fig:cosmic_time} shows the range of plausible threshold times as a function of halo mass $M_0$.
\subsection{Velocity-dependent SIDM cross section} \label{sec_cross section}
The cross section per unit mass, $\sigma/m_\chi$, is the key physical property that characterizes a specific SIDM model. We consider a class of models where self-scattering between DM particles is mediated by a massive force carrier of mass $m_{\phi}$ through an attractive Yukawa potential with coupling strength $\alpha_c$ \cite{Vogelsberger, Feng, Loeb2011}. 
Furthermore, we assume that the elastic scattering between SIDM particles can be modeled by the screened Coulomb scattering interaction observed in a plasma, which in the classical regime is well fitted by the transfer cross section:
\begin{equation}
\dfrac{\sigma_{\rm T}}{\sigma_{\rm T}^{\rm max}} \approx 
     \begin{cases}
       \dfrac{4\pi}{22.7}\beta^2\ln\left(1+\beta^{-1}\right),                    &\beta<0.1\\[15pt]
       \dfrac{8\pi}{22.7}\beta^2\left(1+1.5 \beta^{1.65} \right)^{-1},                   &0.1<\beta<10^3\\[15pt]
       \dfrac{\pi}{22.7}\left(\ln\beta+1- 0.5\ln^{-1}\beta\right)^2,  &\beta>10^3, 
     \end{cases}
\label{eq:cross}
\end{equation}
where $\beta=\pi v_{\rm max}^2/v^2=2\,\alpha_c\,m_{\phi}/(m_{\chi}v^2)$ and $\sigma_{\rm T}^{\rm max}=22.7/m_{\phi}^2$, and $v$ is the
relative velocity of the DM particles. Here $v_{\rm max}$ is the velocity at which $(\sigma_{\rm T} v)$ peaks at a transfer cross
section equal to $\sigma_{\rm T}^{\rm max}$. 

In this work, we use the benchmark velocity-dependent SIDM model introduced in \cite{Zavala2019} with $v_{\rm max}$ = 25 km/s and $\sigma_{\rm T}^{\rm max} = {\rm 60 cm^2g^{-1}}$. Ref.~\cite{Zavala2019} used SIDM cosmological zoom simulations of a MW-size halo to show that self-interactions are frequent enough in the center of dwarf-scale (sub)halos to trigger the gravothermal catastrophe phase in a fraction of the subhalo population, and thus constitutes an alternative explanation to the diverse distribution of inner DM densities found in the MW satellite population (see Section~\ref{sec_intro}).

In our idealised model, we are interested in the characteristic scales of a given halo that are relevant to set a single characteristic cusp-core-collapse timescale. In particular, we assign a single relaxation timescale for a halo due to self-scattering using Eq.~\ref{t_relax}. We set a single characteristic velocity dispersion $\sigma_{\rm vel}$, which is given by the the maximum of the velocity dispersion profile $\sigma_{\rm vel}=\sigma_{\rm r}(r_{\rm max})$ of the primordial NFW halo at the beginning ($z_{\rm cc}$) of the cusp-core transformation. The radius at which this maximum occurs is of $\mathcal{O}(1)$ of the maximum size of the SIDM core that eventually develops, and the value of $\sigma_{r}(r_{\rm max})$ sets the {\it temperature} of the fully developed isothermal core. We now describe in detail how we calculate $\sigma_{r}(r_{\rm max})$.

We start by referring to the local radial velocity dispersion $\sigma_{\rm r}(r)$, which can in principle be obtained self-consistently by solving the Jeans equation:
\begin{equation}    \label{eq:Jeans}
    \dfrac{1}{\rho} \dfrac{\rm d}{{\rm d} r} (\rho \sigma_{\rm r}^2) +
    2 \beta \dfrac{\sigma_{\rm r}^2}{r} = -\dfrac{{\rm d} \Phi}{{\rm d} r} \ ,
\end{equation}
where $\beta=1-\sigma_\theta^2/\sigma_{\rm r}^2$ is the velocity anisotropy parameter and $\Phi$ is the gravitational potential, which for the NFW profile is given by:
\begin{equation}    \label{eq:grav-potential}
    \dfrac{\Phi(s)}{V_{200}^2} = -\dfrac{1}{k(c)}\dfrac{\ln (1 + cs)}{s} \ ,
\end{equation}
where $s=r/r_{200}$ and $V_{200}$ is the circular velocity at $r=r_{200}$:
\begin{equation} \label{eq:V_200}
V_{200}^2 = G \left({\rm M_{0}}^2 \times \dfrac{4}{3} \pi 200 \rho_{\rm crit}\right)^{1/3} .
\end{equation}
Here we assume the simplest case of isotropic orbits, where $\sigma_\theta(r)=\sigma_{\rm r}(r)$ and $\beta=0$. In this case, the solution to the Jeans equation can be computed analytically \cite{Lokas}, giving the 1D velocity dispersion:
\begin{equation}
\begin{split}
    \dfrac{\sigma_{\rm r}^2}{V_{200}^2} (s, \beta=0) 
    = \dfrac{1}{2k(c)}c^2s(1+cs)^2 [\pi^2 - \ln(cs)  - \dfrac{1}{cs} \\
         - \dfrac{1}{(1+cs)^2} -\dfrac{6}{1+cs}+ \left(1 +\dfrac{1}{c^2s^2} - \dfrac{4}{cs} - \dfrac{2}{1+cs} \right) \\
         \times \ln (1+c s) + 3 \ln^2 (1+c s) + 6 {\rm Li}_2(-c s) ],
        \label{eq:sigma1d}
\end{split}
\end{equation}
where ${\rm Li}_2 (x)$ is the dilogarithm. Using Eq.~\ref{eq:sigma1d}, we compute $\sigma_{\rm vel}=\sigma_r(r_{\rm max})$ for a given value of $M_0$ and $c$.

\begin{figure}
    \includegraphics[width=\linewidth, clip=True]{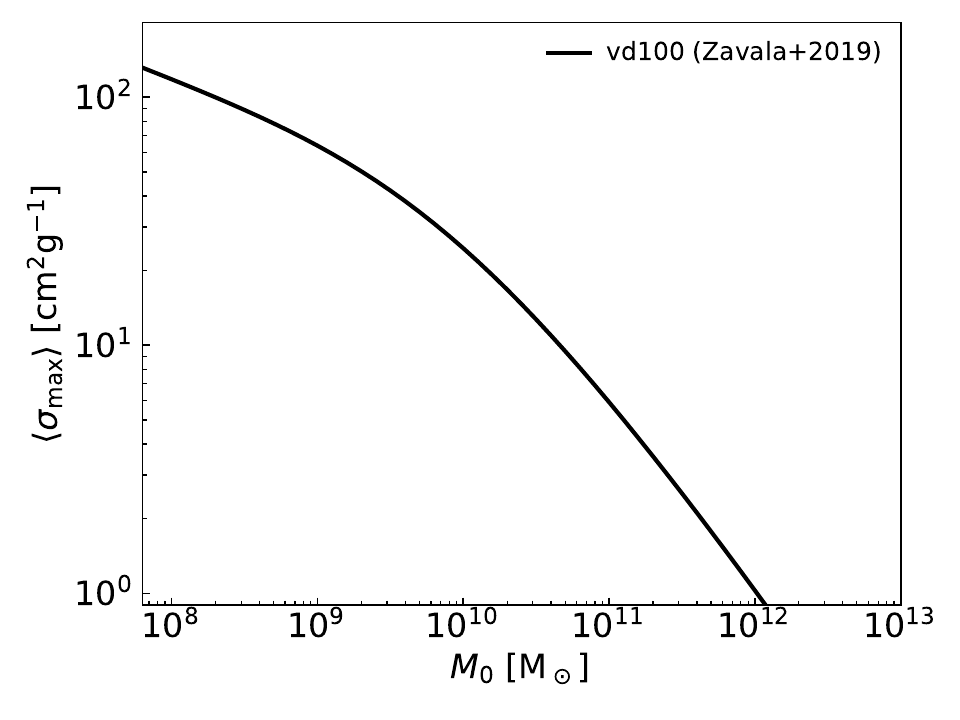}
\caption{Characteristic transfer cross section per unit mass as a function of a halo mass at $z=0$ (Eq.~\ref{eq:avg_cross}) for the benchmark SIDM model (vd100) from \cite{Zavala2019}.}
\label{fig:cross_section}
\end{figure}

Finally, we compute a characteristic value for the transfer cross section $\langle\sigma_{\rm max}\rangle$ by computing the
thermal average of the transfer cross section at $r_{\rm max}$, i.e., within the SIDM core. We assume that the velocity distribution of DM particles can be approximated by a Maxwell-Boltzmann distribution. Although such a distribution is not self-consistent with the NFW profile \cite[e.g.][]{Petac2018}, it is a reasonable approximation for our purposes, because the distribution within the SIDM core will eventually become Maxwellian \cite{Vogelsberger2013}. Therefore: 
%
\begin{equation}
  \langle\sigma_{\rm max}\rangle=\dfrac{1}{2\sigma_{\rm vel}^3\sqrt{\pi}}\int (\sigma_{\rm T}) v^2e^{-v^2/4\sigma_{\rm vel}^2}\,{\rm d}v.
\label{eq:avg_cross} 
\end{equation}
For reference, Fig.~\ref{fig:cross_section} shows the characteristic value for the transfer cross section per unit mass as a function of halo mass $M_0$ today for the benchmark SIDM model (vd100) from \cite{Zavala2019} that we use for calibration in our work.
\subsection{Mass and time scales for black hole formation}\label{sec_BH-formation}
We estimate the timescale for the formation of a black hole in the center of an SIDM halo following the procedure laid out by \cite{Outmezguine2022}. This timescale applies after the threshold time for the cusp-core transformation $z_{\rm cc}$ of the primordial NFW halo as discussed in Section~\ref{sec_redshift}. The formula from \cite{Outmezguine2022} is based on the (spherical) gravothermal fluid model, which has been used in the past to study the gravothermal catastrophe in SIDM halos \cite[e.g.][]{Balberg01,Koda2011,Essig2019,Pollack, Xiao2021, Choquette2019, Feng21}. The recent study by \cite{Outmezguine2022} is particularly suitable for our work since, contrary to previous papers, they consider velocity-dependent cross sections. In Appendix~\ref{sec_vel_independent_t_r} we make a comparison between their estimate of the collapse time and that made in \cite{Pollack}. Before describing the key equation, we present a short summary of the physical process based on \cite{Balberg01}.

In the gravothermal evolution of an SIDM halo, we can distinguish between a long mean free path (LMFP) regime, where the typical distance a particle travels is much longer than the gravitational Jeans scale and the short mean free path (SMFP) regime, where the situation is reversed. In the LMFP regime, particles orbit the inner halo many times unperturbed before being scattered, while in the SMFP regime particle motions in the core are constrained by multiple collisions. In these two regimes, the heat conduction and the mass transfer between the core and the extended halo are different.

Initially, as the core is growing in size, both the core and extended halo are firmly in the LMFP limit; during this phase, the inner core is approaching the transitional regime, in between the LMFP and the SMFP where the Knudsen number is of $\mathcal{O}(1)$. While the extended halo remains nearly collisionless, with nearly the same primordial NFW density profile, the core evolves and, as the gravothermal collapse is triggered, it transitions into the SMFP regime where it essentially behaves as a fluid. Although the velocity dispersion and the density of the core both increase, the latter increases much faster with time,
$\rm d\log(\sigma_{\rm r}^2)/\rm d\log(\rho_{\rm core}) \approx 0.1$ \cite{Balberg01}, 
and it drives the core into two components: a dense SMFP inner core that continues to evolve, and a more dilute LMFP outer core, with a nearly constant density, which connects to the extended nearly unperturbed NFW halo. Eventually, the temperature in the inner core is so high that it enters the relativistic regime and dynamical instability occurs that leads to the formation of the black hole \cite{Balberg01} (see also \cite{Feng2021}).
The {\it classical} gravothermal fluid formalism cannot be used once the system becomes relativistic. However, the classical formalism
in the LMFP regime allows us to follow the evolution of the core to high central densities, and in fact since most of the evolution occurs in the LMFP regime, with only the last instants prior to collapse being in the SMFP regime, the timescale for collapse is dominated by the LMFP evolution of the core, and thus, the classical approach can be used to estimate the timescale of interest.

According to \cite{Outmezguine2022}, the amount of time required for the primordial NFW to evolve from a central cusp to an SIDM core and then to a fully collapsed cored when a black hole forms -- i.e. the cusp-core-collapse timescale -- is given by:
\begin{equation}\label{t_col}
t_{\rm coll} \approx 382t_{\rm r}(z_{\rm cc}). 
\end{equation}
where $t_{\rm r}(z_{\rm cc})$ is the relaxation time, defined as the mean time between individual collisions, which we introduced briefly at the beginning of Section~\ref{cosmic} (see Eq.~\ref{t_relax}). Ref.~\cite{Outmezguine2022} gives a formula for $t_{\rm r}(z_{\rm cc})$ based on the properties of the primordial NFW halo:
\begin{equation}
\begin{split}
t_r(z_{\rm cc}) &\simeq 1.47 {\rm Gyr} \times \left(\dfrac{0.6}{C}\right) \times \\
&\times \left(\dfrac{\rm{cm^2/g}}{\sigma_{\rm c,0}(z_{\rm cc})}\right) \left(\dfrac{100\rm{km/s}}{\sigma_{\rm vel}(z_{\rm cc})}\right) \left(\dfrac{10^{7}\rm{M_{\odot}}\rm{kpc}^{-3}}{\rho_{\rm s}(z_{\rm cc})}\right),
\end{split}
\label{eq:t_rel}
\end{equation}
xwhere $C$ is a fitting parameter, which we set to $0.57$ following \cite{Outmezguine2022}. The parameter $\sigma_{c,0}$ is a type of cross section average given by:
\begin{equation}\label{eq:sig_c}
	\sigma_{c,0}(z_{\rm cc})= \dfrac{3}{2}
	\dfrac{\left<\sigma_{\rm visc} v^3\right>}{\left<v^3\right>},
\end{equation}
 where $\sigma_{\rm visc}=\int \rm d \sigma \sin^2 \theta$ is the viscosity cross section with the scattering deflecting angle $\theta$. The difference between $\sigma_{\rm visc}$ and $\sigma_{\rm T}$ is small and depends on the SIDM particle physics model, which gives the specific angular dependence for the differential cross section \cite{Tulin2018}. We assume that $\sigma_{\rm visc}\approx\sigma_{\rm T}$ for isotropic scattering, since our results are calibrated on the SIDM simulations by \cite{Zavala2019}, which use elastic isotropic scattering using $\sigma_{\rm T}$.
 The cosmic time from the Big Bang until the formation of the black hole is finally given by:
\begin{equation}
t_{\rm BH}=t(z_{\rm cc})+382t_r(z_{\rm cc}),
\label{eq:collapse_time}
\end{equation} 
where $t(z_{\rm cc})$ is the time at which the core-collapse transition begins, and is computed with Eq.~\ref{eq:cosmic_time}.

The black hole is expected to form from material in the SMFP region. 
Ref.~\cite{Balberg02} estimates the mass of this seed black hole $M_{\rm BH}$ based on the mass in the core, $M_{\rm core}$, that is in the SMFP regime. In the late stages of the core evolution, the $M_{\rm core}$ - $\sigma_{\rm vel}$ relation determines the mass of the core at the relativistic instability, which occurs when $\sigma_{\rm vel} \approx c/3$.  When the inner core is sufficiently dense, mass is continuously lost from its surface as outer layers cool and expand to join the outer core, with \cite{Balberg01} predicting that the $M_{\rm core}$ - $\sigma_{\rm vel}$ relation settles to: $\rm d\log M_{\rm core}/\rm d\log(\sigma_{\rm vel}^2)\approx -0.85$. Therefore, the seed black hole mass is predicted to be:
\begin{equation}\label{eq:M_bh}
    M_{\rm BH} (z^\ast) = M_{\rm core}(z^\ast) \left(\dfrac{\sigma_{\rm vel }^2(z^\ast)}{(c/3)^2\rm {km^2 s^{-2}}}\right)^{0.85}. 
\end{equation}
The region outside the collapsed core relaxes to a dynamically stable equilibrium system of particles that continue to orbit the central black hole and subsequent interactions in this region will feed the black hole.

We should note that the behaviour above has been developed for a system in isolation. In an evolving halo growing in a cosmological scenario, mass accretion might modify this behaviour as we described in Section~\ref{sec_redshift}. A detailed treatment of the impact of cosmological accretion in the gravothermal fluid equations, and in particular on the scale of the SMFP region, goes beyond the scope of this work. Instead, we consider a simple approach in which we establish a range of plausible black hole masses by considering the epoch $z^\ast$ at which Eq.~\ref{eq:M_bh} should be evaluated. A lower estimate for the seed black hole mass would be to set $z^\ast=z_{\rm cc}$, that is to assume that the scale of the core (LMFP region), and thus the scale of the collapsing SMFP region, is set by the properties of the halo, essentially its mass $M(z_{\rm cc}$), at the threshold time for the cusp-core transformation. This, however, ignores the fact that the halo mass grows during the cusp-core transformation and up to the point of collapse ($z_{\rm BH}$ given by Eq.~\ref{eq:collapse_time}), the size of the core should thus grow as well, affecting the scale of the collapsing region. An upper estimate for the seed black hole mass can then be given by setting $z^\ast=z_{\rm BH}$, that is to assume that the size of the SMFP region is set by the last stages of the cusp-core-collapse evolution when the LMFP region (core) has grown to a size set by $M(z_{\rm BH})$. We then bracket the plausible range of black hole masses by:
\begin{equation}\label{eq:m_bh_range}
M_{\rm BH}(z^\ast=z_{\rm cc})<M_{\rm BH}<M_{\rm BH}(z^\ast=z_{\rm BH}).
\end{equation}
To estimate the core mass $M_{\rm core}$ in Eq.~\ref{eq:M_bh}, we use the results of the gravothermal fluid approach in \cite{Outmezguine2022} (the same reference we use in Eq.~\ref{eq:t_rel}) where is estimated that the maximal core size of the halo is:
\begin{equation}\label{eq:r_core}
    r_{\rm core}\simeq 0.45 r_{-2},
\end{equation}
before the collapse regime begins. This core radius is defined as the radius at which the local density is half that of the central density. The core size estimate in Eq.~\ref{eq:r_core} is roughly consistent with simulation results where the core size is found to be $\lesssim r_{-2}$ \cite{Rocha2012,Vogelsberger2014}. Assuming that the region beyond the core remains essentially collisionless and with a profile that matches the NFW distribution we then have:
\begin{equation}
    M_{\rm core}(z^\ast)=M_{\rm NFW}(r_{\rm core};z^\ast),
\end{equation}
where $M_{\rm NFW}(r)$ is given by the NFW radial density profile (Eq.~\ref{eq:NFW}) for a halo with a mass $M_0(z^\ast)$ given by the mass accretion history formula (Eq.~\ref{eq:M_mean}) evaluated at $z^\ast$, and a concentration $c(z^\ast)$ obtained from this mass using the model described in Section~\ref{sec_CM-relation}.

\subsection{Impact of tidal stripping in the core-collapse phase}\label{sec_tidal-stripping}

Tidal stripping is the process by which DM in the outskirts of a smaller halo is removed by tidal forces as it orbits within a larger host.
It has been argued that the tidal interactions with the halo and central galaxy of the host accelerate the core-collapse timescale by increasing the {\it temperature} (velocity dispersion) gradient outside the core, making heat outflow more efficient \cite{Kaplinghat2019, Nishikawa, Kahlhoefer2019, Sameie2020}. These previous works have invoked tidal acceleration of core-collapse as an explanation for the diversity of the MW's dwarf spheroidal galaxies (dSphs), based on constant cross section SIDM models with relatively low cross sections $\sigma_{\rm T}/m_\chi\sim1-5$~cm$^2$g$^{-1}$, while the velocity-dependent model presented in \cite{Zavala2019} \cite[see also][]{Correa2021,Turner} relies on large cross sections $\sigma_{\rm T}/m_\chi>10$~cm$^2$g$^{-1}$ at the characteristic velocities of the dSphs to ensure core-collapse.

In order to take into account the impact of tidal stripping in accelerating the core-collapse phase, we use the results presented in \cite{Nishikawa}, where tidal stripping is assumed to modify the NFW profile for $r>r_t$ in the following way: $\rho_{\rm NFW}(r_t)\times(r_t/r)^{p_t}$ where $p_t$ = 5 (based on \cite{Penarrubia}) and $r_t$ is a truncation radius. This modified profile is a simple way to incorporate the impact of mass loss from the outer region in the timescale for collapse. 
We use the case of $r_t = r_{-2}$, and estimate the acceleration of the gravothermal catastrophe due to tidal stripping as:
\begin{equation}\label{t_col_tidal}
    t_{\rm coll,t}\approx\dfrac{1}{10}t_{\rm coll},
\end{equation}
where $t_{\rm coll}$ is the timescale for core-collapse without the tidal effects (see Eq.~\ref{t_col}). We only apply Eq.~\ref{t_col_tidal} for (sub)halos after the infall redshift into the MW halo, which for simplicity, we fix to be $z_{\rm infall}=1$ for the entire (sub)halo mass range we consider in this work\footnote{We note that although the subhalo infall redshift distribution is broad, depending on the mass and orbit of individual subhalos, it roughly has a median value of $z\sim1$ for the subhalo population of MW-size halos \cite{Rocha2012,Jian2015}. Since this is the population we are interested in, and since we are not considering detailed orbital properties, we fix $z_{\rm infall}=1$.}.

\begin{figure*}
    \includegraphics[width=1\textwidth,  clip=true]{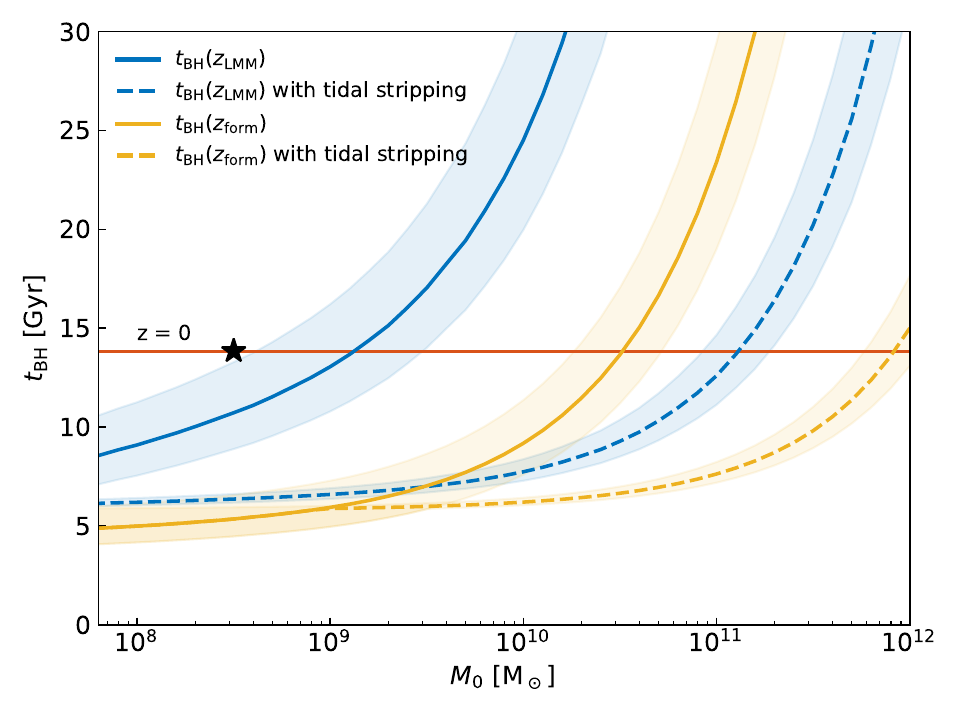}
\caption{The timescale to form a black hole (Eq.~\ref{eq:collapse_time}) in SIDM halos as a function of the present-day halo mass $M_0$ (in isolation) for the velocity-dependent model vd100 \cite{Zavala2019,Turner} shown in Fig.~\ref{fig:cross_section}. The yellow solid line shows the case where the starting time for the cusp-core-collapse evolution is set to the assembly/formation time of the primordial (pre-SIDM) CDM NFW halo: $z_{\rm cc}=z_{\rm form}$ (see Section~\ref{sec_CM-relation} and Fig.~\ref{fig:cosmic_time} where the yellow dotted line marks the cosmic time corresponding to $z_{\rm form}$), while the blue solid line brackets the impact of cosmological accretion by setting $z_{\rm cc}=z_{\rm LMM}$, which is the epoch of the last major merger for given halo of mass $M_0$ (see Section~\ref{sec_redshift} and Fig.~\ref{fig:cosmic_time} where the blue dotted line marks the cosmic time corresponding to $z_{\rm LMM}$).
Shaded regions indicate a scatter of $\pm10\%$ in the concentration-mass relation \cite[e.g.][]{SC2014}.
Dashed lines are the corresponding cases including the acceleration of the collapse time driven by tidal stripping \cite{Nishikawa} assuming a mass-independent infall epoch of $z_{\rm infall}=1$ (see Section~\ref{sec_tidal-stripping}). 
The horizontal orange line indicates the age of the Universe. The black star symbol marks the transition mass where 50\% of the subhalo population is estimated to be in the core-collapse regime according to \cite{Lovell2022}; see also \cite{Zavala2019,Turner}.}
\label{fig:formation_time}
\end{figure*}

\section{Results and Discussion}\label{sec_results}
Our main goal is to investigate the consequences for velocity-dependent SIDM models, which invoke core-collapse to explain the diversity of the MW satellite population (such as that in \cite{Zavala2019}), on the formation timescales for IMBHs and their masses.

Fig.~\ref{fig:formation_time} shows the gravothermal collapse timescale -- which is approximately the BH formation timescale -- as a function of the halo's present day mass $M_0$ (Eq.~\ref{eq:collapse_time}). The solid lines -- together with their shaded regions, which are given by the scatter in the cosmological concentration--mass relation -- bracket the range of possible threshold epochs for the cusp-core-collapse evolution to begin (see Section~\ref{sec_redshift}), with the yellow corresponding to the assembly redshift of the primordial NFW halo $z_{\rm cc}=z_{\rm form}=z_{-2}$, while the blue corresponds to the epoch of the last major merger $z_{\rm cc}=z_{\rm LMM}$. 
The points where the right edge of the blue region, $t_{\rm BH}(z_{\rm LMM})$, and the right edge of the yellow region, $t_{\rm BH}(z_{\rm form})$, cross the $z=0$ horizontal line roughly indicate the maximum mass of an isolated SIDM halo that could undergo core-collapse by the present day, for both of these cases. 
For the case where $z_{\rm cc}=z_{\rm LMM}$, the maximum mass is $\sim 3 \times 10^9 \rm M_\odot$ and for $z_{\rm cc}=z_{\rm form}$, it is  $\sim 5 \times 10^{10} \rm M_\odot$.

Since we are interested in the (sub)halos that could host the MW satellites, we have considered the impact of tidal effects in the timescale for black hole formation. As discussed in Section~\ref{sec_tidal-stripping}, for simplicity we assume that all (sub)halos in the mass range considered in Fig.~\ref{fig:formation_time} become satellites at an infall redshift $z_{\rm infall}=1$, and that tidal forces by the host halo strip the material from the subhalo making the heat outflow from the center to the outskirts of the subhalo much more efficient, and thus reduces the timescale for collapse by a factor of ten (see Eq.~\ref{t_col_tidal}). 
This significant acceleration of the core-collapse phase is shown as dashed lines in Fig.~\ref{fig:formation_time}, which shifts the upper limit of the mean mass of a halo that could undergo the core-collapse to $\sim 2 \times 10^{11} \rm M_\odot$ and $\sim 10^{12} \rm M_\odot$ for the $t_{\rm BH}(z_{\rm LMM})$ and $t_{\rm BH}(z_{\rm form})$ cases, respectively. 

The black star symbol in Fig.~\ref{fig:formation_time} indicates the halo mass at which 50\% of (sub)halos are estimated to undergo core-collapse by $z=0$ according to the results in Ref. ~\cite{Lovell2022}, which is based on the simulation of the vd100 model presented in \cite{Zavala2019}. The simulation results are not compatible with our modelling of an early cusp-core transformation (yellow lines), and/or the acceleration effect due to tidal stripping (dashed lines). In the following, we discuss the effects that are likely behind this result.

{\it Tidal acceleration of core-collapse?} 
The impact of additional environmental effects taking place between a (sub)halo and the host halo during mergers, such as the evaporation of subhalo particles due to interactions with particles in the host, have been found to counteract the tidal stripping effect, delaying -- or even disrupting -- the core-collapse phase in models with low cross sections $\sigma_{\rm T}/m_\chi\lesssim10$~cm$^2$g$^{-1}$ \cite{Zeng2022}. Recent $N-$body cosmological SIDM simulations of a MW-size halo and its subhalos with a cross section in the range $\sigma_{\rm T}/m_\chi\sim1-5$~cm$^2$g$^{-1}$ confirm that subhalos do not experience core-collapse, thus larger values are required \cite{Silverman}. These recent results essentially rule out the constant cross section SIDM model as a viable possibility to explain the diversity of the MW satellite population, and therefore strengthen the case for a velocity-dependent SIDM model with core-collapse such as the one explored here based in \cite{Zavala2019} (see also \cite{Correa2021}). Moreover, these results indicate that core-collapse is not accelerated in the manner anticipated by the tidal stripping model in \cite{Nishikawa}, and thus the dashed lines in Fig.~\ref{fig:formation_time} are likely overestimating its effectiveness.

 \begin{figure}
    \includegraphics[width=\linewidth, clip=True]{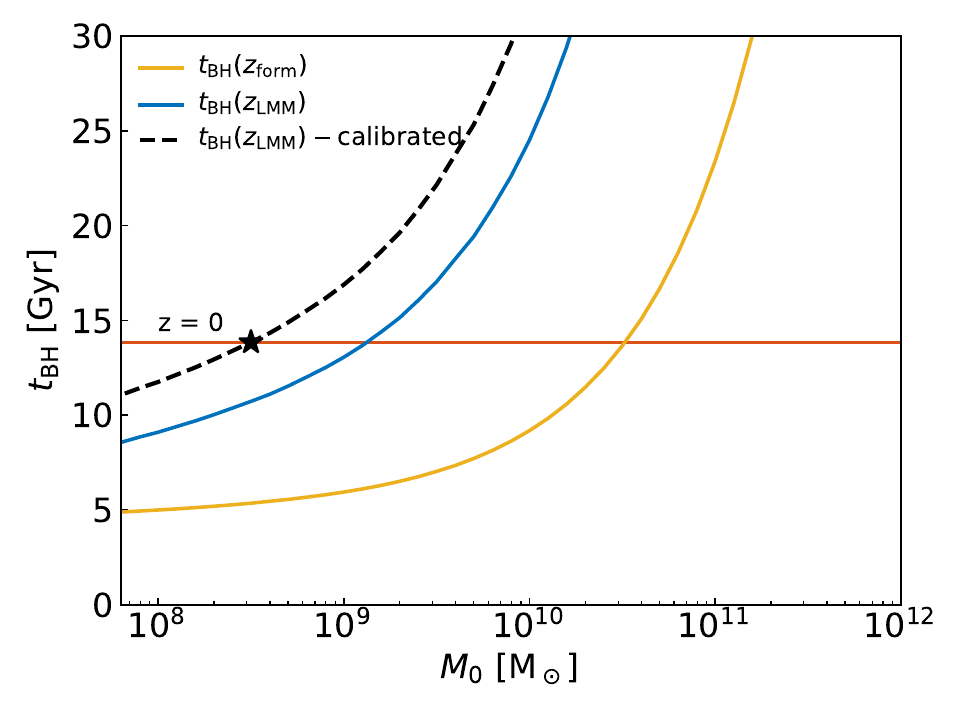}
\caption{Timescale for black hole formation in SIDM halos as a function of the present-day SIDM halo mass $M_0$ in isolation. The solid blue and yellow lines are the same as those in Fig.~\ref{fig:formation_time}, which bracket the range of the possible threshold epochs for the cusp-core-collapse evolution to begin (for mean values of the concentration-mass relation). The dashed black line is $t_{\rm BH} (z_{\rm LMM})$ (i.e., the blue line) re-calibrated to the simulation-based result of Ref.~\cite{Lovell2022} (black star) with the calibration factor $C=0.42$ (see Eq.~\ref{eq:t_rel}).}
\label{fig:viable-mass-range}
\end{figure}
{\it Cosmological accretion.} Based on the previous discussion, the likely range of validity for our modelling is shown in Fig.~\ref{fig:viable-mass-range}, where we have omitted the concentration--mass relation scatter for clarity. It is clear that the simulation result (black star symbol) is closer to the model where the cusp-core transformation begins later, at $z_{\rm cc}=z_{\rm LMM}$. This finding supports the case for cosmological accretion increasing the time of the evolution of a SIDM halo spent in the cored, quasi stable regime, possibly due to energy injection from infall material into the central core as discussed in Section~\ref{sec_redshift}. A good match to the simulation results can then be achieved by setting $z_{\rm cc}=z_{\rm LMM}$ in our model and adjusting the $C$ parameter in Eq.~\ref{eq:t_rel} to $C\approx0.42$, which is shown as a dashed line in Fig.~\ref{fig:viable-mass-range}. We note that $C$ is a calibration factor which, following \cite{Outmezguine2022}, we had set to $C=0.57$ in Fig.~\ref{fig:formation_time}. This agrees with \cite{Essig2019} who calibrated this parameter to a very similar value using the {\it isolated} SIDM simulations in Ref.~\cite{Koda2011}. We can then model the impact of cosmological accretion by either setting $z_{\rm cc}$ to $z_{\rm LMM}$ and make a small modification to $C$ (which is the case we adopt), or by modifying the value of $C$ significantly (starting from $z_{\rm cc}=z_{\rm form}$) and invoking a needed re-calibration of the parameter based on cosmological simulations. For our model, the latter case can be achieved by setting $z_{\rm cc}=z_{-2}$ and fixing $C\approx0.19$. 

\subsection{IMBHs in the ultra-faint galaxies}

\begin{figure}
    \includegraphics[width=\linewidth, clip=True]{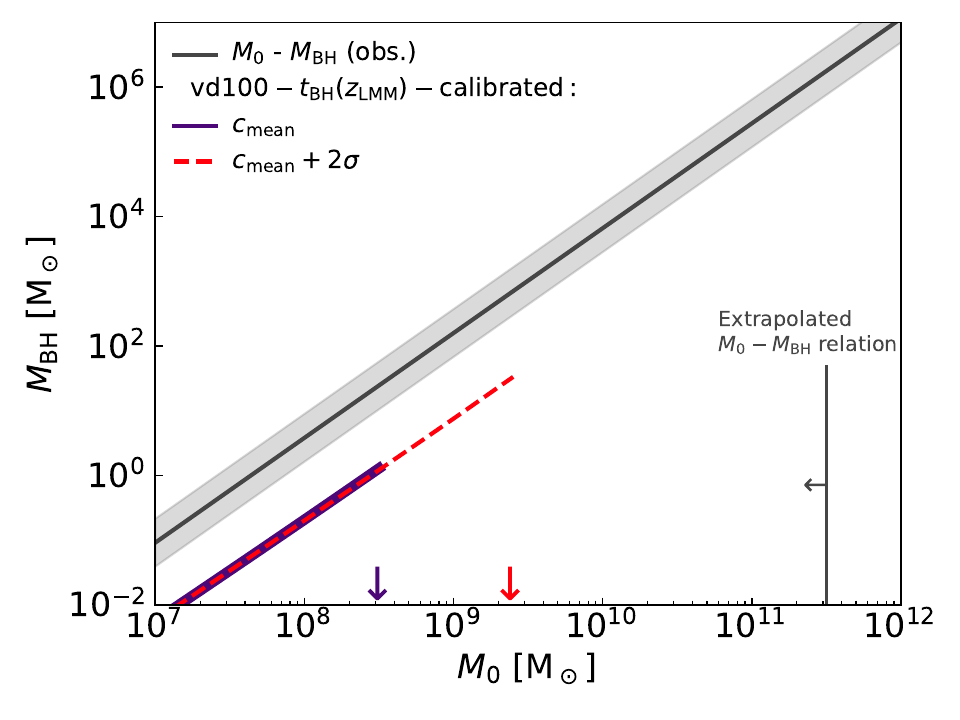}
\caption{Black hole mass--halo mass relation. The violet and red lines follow the estimate of the SIDM-core-collapse formula (Eq.~\ref{eq:M_bh}) with the former using the mean concentration-mass relation, while the latter uses a $+2\sigma$ value over the mean. For these cases, the arrows indicate the corresponding halo mass at which $t_{\rm BH}=0$, i.e., the mass at which 50\% and 2.5\% of the halo population at that mass is in the core-collapse regime, violet and red, respectively. The gray line is the extrapolation towards lower masses of the empirical relation for SMBHs in galaxies with halo masses $>10^{11.5} \rm{M_\odot}$, while the gray band represents the intrinsic scatter on this relation; adopted from \cite{Marasco2021}. }
\label{fig:M0-Mbh}
\end{figure}

One consequence of invoking core-collapse of SIDM halos to explain the diversity of inner DM densities in the MW satellite population is that those collapsed satellites will host central black holes. In particular, the vd100 model explored in \cite{Zavala2019} predicts that the dense ultra-faint galaxies, specifically, Segue~I, Segue~II, Willman~I and Boötes~II will be hosted by gravothermally collapsed subhalos (see Fig.~3 in \cite{Zavala2019}). We can compute the expected seed black hole mass for a collapsed SIDM halo in the vd100 model using our framework (see Eq.~\ref{eq:M_bh}). This is shown in Fig.~\ref{fig:M0-Mbh} as a function of $M_0$, the halo mass in isolation. We use our calibrated model with $z_{\rm{cc}}=z_{\rm{LMM}}$ and $C=0.42$ and consider two cases to illustrate the impact of halo concentration in our results: i) the solid violet line where halos have a mean concentration at $M_0$ ($c_{\rm mean}$) and ii) the dashed red line for halos with a concentration in excess\footnote{Recall we are using a mass-independent spread of the distribution of halos in the concentration-mass relation equivalent to 0.1~dex for $1\sigma$ of the distribution.} to the mean by $2\sigma$ ($c_{\rm mean}+2\sigma$). We first notice that although the two lines representing these cases almost overlap with each other in Fig.~\ref{fig:M0-Mbh}, they in fact have a different slope and normalization since Eq.~\ref{eq:M_bh} depends (weakly) on concentration\footnote{This dependence is not strong because $z_{\rm BH}$ decreases with $M_0$, thus the halo mass at this redshift, $M(z_{\rm BH})$, increases with $M_0$, which makes the relevant concentration, $c(M(z_{\rm BH}))$, almost independent of present day halo mass $M_0$.}. The net impact of concentration in the value of $M_{\rm BH}$ for a given halo mass is up to 5\%.

Secondly, these predicted $M_{\rm BH}-M_0$ relations are truncated at different halo masses, represented by the vertical downward arrows of the respective color. In the first case (using $c_{\rm mean}$), this cutoff mass occurs at $M_0\sim3\times10^8$~M$_\odot$ (violet arrow), and can be interpreted as the mass at which 50\% of the halos have core-collapsed and 50\% of the halos are still in the core phase. At higher (lower) masses, the fraction of core-collapsed halos is lower (higher) depending on halo concentration. For example, SIDM halos with $M_0\sim2\times10^9$~M$_\odot$ are only expected to have collapsed by the present day if they have a concentration larger than the mean by $2\sigma$ (red arrow), which represents only a small fraction ($2.5\%$) of the halos at this mass. Therefore, in the vd100 model, only a small fraction of the massive (sub)halos in a MW-size system, which are expected to host the MW satellites, would have undergone core-collapse.

As we discussed in Section~\ref{sec_BH-formation}, the range of plausible BH masses depends on the epoch ($z^\ast$) at which one should evaluate the relevant properties of the core that determine the SMFP region that collapses (Eq.~\ref{eq:m_bh_range}). We established $z^\ast=z_{\rm BH}$ as a reasonable choice and it is the one that appears in Fig.~\ref{fig:M0-Mbh} and in subsequent figures. Modifying this choice to the earliest plausible epoch $z^\ast=z_{\rm cc}$ results instead in a smaller BH mass. For our default choice of $z_{\rm cc}=z_{\rm lmm}$ the difference is up to 30\% over the relevant mass range with a weak dependence on halo mass.

The gray line in Fig.~\ref{fig:M0-Mbh} represents the observed supermassive black hole (SMBH)-mass--halo-mass relation for massive galaxies extrapolated to low masses \cite{Marasco2021}, while the shaded gray band represents the intrinsic observational scatter. This relation and its spread have only been measured in halos larger than $10^{11.5}$~$\rm{M_\odot}$, and therefore we plot the extrapolated values down to $10^{7}$~$\rm{M_{\odot}}$ in Fig.~\ref{fig:M0-Mbh}. Such an extrapolation of the empirical $M_{\rm BH}-M_0$ relation to the regime of dSphs ($M_0\leq10^{10}\rm{M_\odot}$) would imply IMBHs in the range between $4-7\times 10^3$~M$_\odot$. Remarkably, the slope of the predicted SIDM-driven $M_{\rm BH}-M_0$ relation by 
(Eq.~\ref{eq:M_bh}) is very similar to that of the SMBH--halo mass relation, while the normalization is approximately two orders of magnitude lower.

\begin{figure*}
    \centering
    \includegraphics[width=1\textwidth, trim={0cm 0cm 1cm 0cm}]{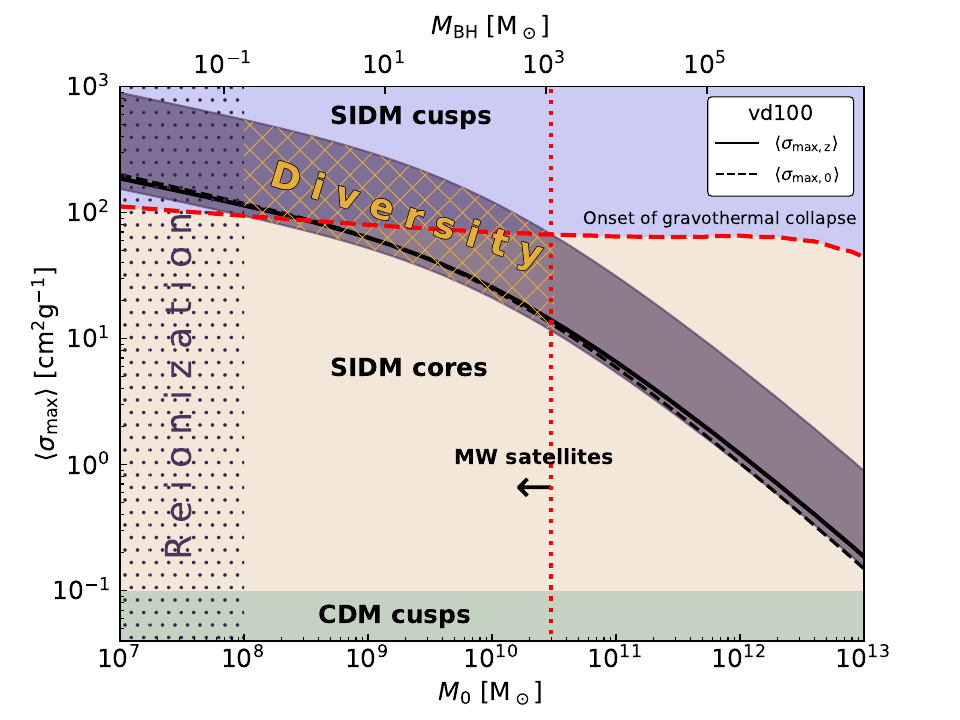}
    \caption{ The effective cross section as a function of a present-day SIDM halo mass $M_0$ in isolation. The black solid and dashed lines represent the effective cross section for the vd100 model evaluated at $z_{\rm LMM}$ and $z=0$, respectively; we use our default model with $z_{\rm cc}=z_{\rm LMM}$ and calibrated to the simulation analysis in \cite{Lovell2022}. The x$-$axis on the top shows the corresponding (SIDM-driven) black hole mass for a given $M_0$. The red dashed line (nearly horizontal) indicates the required cross section value for the onset of gravothermal collapse: SIDM-driven cuspy halos lie above (light violet), while SIDM cores lie below (beige) down to the point where the cross section is so low that halos are essentially CDM-like (light green). The red dotted line marks the upper mass for the dSph MW satellites to reside \cite[e.g.][]{Errani2022}, while the hashed/dotted region to the left starts at the mass where reionization significantly suppresses galaxy formation. The dark violet band indicates the region where vdSIDM models like vd100 but with different normalization, produce a diverse MW satellite population hosted by halos that could either be cored or cuspy.}
    \label{fig:sigma-M}
\end{figure*}

An important aspect of our model is that it aims to explain the diversity of MW satellite population, by invoking a velocity-dependent SIDM model. This model predicts that only a fraction of the satellites have undergone collapse, and therefore only a fraction host IMBHs, specifically the least massive (in a model like vd100, those with $M_0\lesssim3\times10^9\rm{M_\odot}$). For instance, the massive central black hole inferred recently in the dSph Leo~I with $M_{\rm BH}\sim3\times10^6$M$_\odot$ \cite{Bustamente2021} is too massive to lie in the extrapolated $M_{\rm BH}-M_0$ relation, and the properties of the halo associated with Leo~I, being one of the brightest satellites of the MW, would likely put it in a range of $M_0$ values close to, but nevertheless above, the threshold for collapse in the vd100 model. The significance of this issue depends on the specific velocity-dependent SIDM model assumed. For instance, in \cite{Correa2021}, Leo~I is associated with a halo of an initial mass of $M_{200}\sim3\times10^9\rm{M_\odot}$ that has gravothermally collapsed, according to the inferred velocity-dependent SIDM model tuned to explain the diversity of the dSph satellite population in that work. Nevertheless, the central black hole mass inferred for Leo~I in \cite{Bustamente2021} is several orders of magnitude larger than can be explained by gravothermal collapse alone and a significant growth of the seed black hole by other means would be required. For other bright dSphs, there might be a different type of conflict; for instance, Fornax and Ursa Minor have upper limits to the presence of a central black hole of around $M_{\rm BH}\sim3\times10^4\rm{M_\odot}$ \cite{Jardel2012,Lora2009}. These systems have $M_0$ values likely in the range around the threshold for core-collapse. The precise value of $M_0$ inferred from the observed kinematics of the dSph depends on several quantities, such as the modelled DM profile and the orbital parameters, but being both associated to cored systems in \cite{Zavala2019} for the vd100 model, they are not expected to be associated with collapsed subhalos.

We can classify SIDM models in the cross section--halo mass parameter space as to whether they generate a combination of cored and gravothermally collapsed halos in the dwarf galaxy regime. More specifically, we determine the normalization boundaries of the self-interacting cross section according to whether or not the gravothermal collapse regime is expected to occur in a fraction of the MW satellites' host subhalos. In practice, and for simplicity, we only consider SIDM models described by the classical velocity-dependent formula for a Yukawa-like interaction model $\sigma_{\rm T}$ -- i.e. Eq.~\ref{eq:cross} -- and, at first, we fix the relative velocity $v_{\rm max}$ at which $(\sigma_{\rm T} v)$ peaks; in this way, we only vary the normalization $\sigma_{\rm T}^{\rm max}$. 

Fig.~\ref{fig:sigma-M} shows the effective cross section $\langle\sigma_{\rm max}\rangle$ (thermal average of $\sigma_{\rm T}/m_\chi$ at $r_{\rm max}$; see Eq.~\ref{eq:avg_cross}) as a function of a halo mass at the present day, $M_0$. The corresponding BH masses for a given $M_0$ are plotted on the top x$-$axis using Eq.~\ref{eq:M_bh}. The vd100 model results for the cases in which $\langle\sigma_{\rm max}\rangle$ is computed at $z=z_{\rm cc}=z_{\rm LMM}(M_0)$ and at $z=0$ are shown as the black solid and dashed lines, respectively. The former is the relevant cross section to set the core-collapse timescale -- notice that the relevant epoch is a function of mass -- while the latter is shown as reference to make the connection with Fig.~\ref{fig:cross_section} where the thermal average is evaluated at $z=0$ for all halos.

The hashed/dotted region at $M_0\leq 10^8$\msun marks an approximate lower limit on the mass of halos where galaxy formation is efficient; below this mass, heating during the epoch of reionization severely reduces the efficiency of cooling and subsequent star formation \cite[e.g.][]{Sawala2016}. The vertical dotted red line is an approximate upper limit on the mass of possible halo hosts for the population of dSph MW satellites \cite[e.g.][]{Errani}. Thus, the mass range $10^8 - 3\times 10^{10}$\msun represents the region inhabited by the MW satellites. Different colored regions indicate the range of cross section values where halos with different inner density profiles reside. 
The light-green region at $<0.1$~cm$^2$g$^{-1}$ is where DM is effectively collisionless, and thus the structure of all halos is indistinguishable from CDM (i.e. cuspy)\footnote{The upper boundary here is approximate since small cores are expected even at such low cross sections; however, simulation results \cite[e.g.][]{Zavala2013} have shown that these cores are too small at the scale of MW satellites to constitute a significant deviation from the CDM case (see also \cite{Zavala2019review}).}. The light brown region in the middle is where SIDM models deviate significantly from CDM and predict quasi-equilibrium cored halos. The red dashed line marks the effective cross section 
for the onset of gravothermal collapse to occur by $z=0$; it demarcates the transition from cores to SIDM-driven core-collapsed (cuspy) halos (light violet region on the top of Fig.~\ref{fig:sigma-M}).

The violet band in Fig.~\ref{fig:sigma-M} denotes the set of cross section normalization values ($\sigma_{\rm T}^{\rm max}$ in Eq.~\ref{eq:cross}) in vd100-like models that generate a 
diverse population of MW satellites, i.e., where halos with cored and cuspy profiles coexist. This section of parameter space is highlighted with a yellow hatched region 
within the range of halo masses expected for the MW satellites. 
Given that the timescale for core-collapse depends on halo mass and concentration, we can estimate the fraction of halos of a given mass $M_0$ that have undergone gravothermal collapse by considering the probability distribution (PDF) of concentrations for halos at a fixed mass, which according to simulations follows a log-normal distribution \cite[e.g.][]{Neto2007}:
\begin{equation}
P(\log_{10}c)=\dfrac{1}{\sigma
\sqrt{2\pi}}\exp{\Bigl[-\dfrac{1}{2}\Bigl(\dfrac{\log_{10}c-
\langle\log_{10}c\rangle}
{\sigma}\Bigr)^2} \Bigr].
\label{eq:lognormal}
\end{equation}
where $\langle {\rm log}_{10} c\rangle$ is the median value of the concentration (in logarithm) and $\sigma$ is its standard deviation. In our framework, the former is given by the concentration--mass relation model described in Section~\ref{sec_CM-relation}, while the latter is taken to be mass-independent and fixed to 0.1 dex based on simulations.

\begin{figure}
    \includegraphics[width=\linewidth, clip=True]{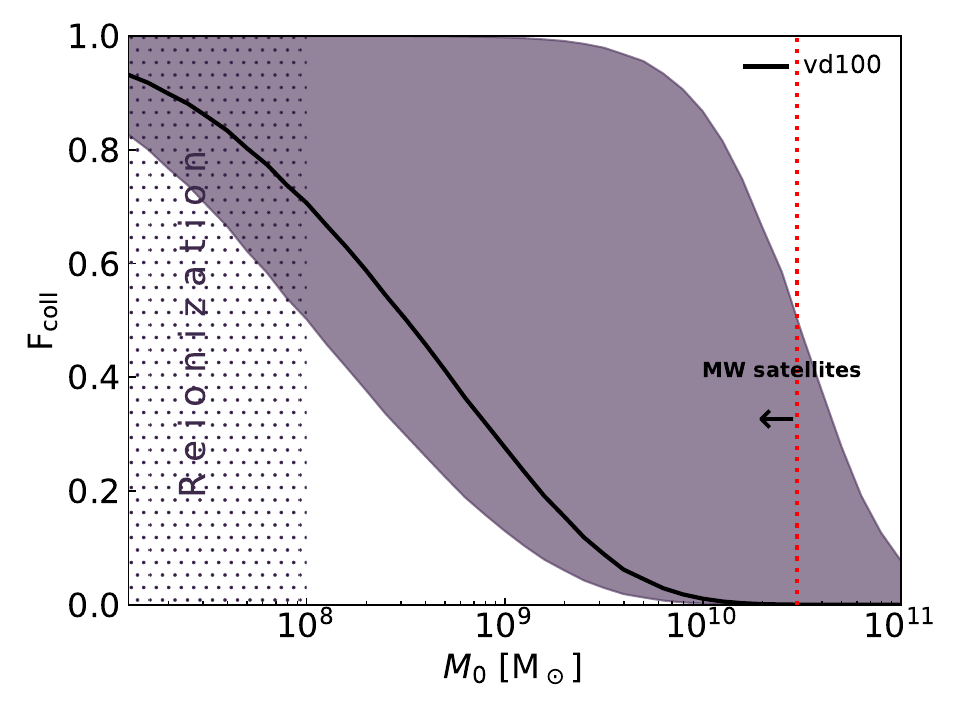}
\caption{The fraction of core-collapsed halos as a function of $M_0$ for the vdSIDM models shown in Fig.~\ref{fig:sigma-M}, which have a diverse range of halo profiles (dark violet). The black solid line represents the benchmark vd100 model. As in Fig.~\ref{fig:sigma-M}, we use our default model with $z_{\rm cc}=z_{\rm LMM}$ and calibrated to the simulation analysis in \cite{Lovell2022}.} 
\label{fig:fcoll}
\end{figure}

Fig.~\ref{fig:fcoll} shows the fraction of core-collapsed halos as a function of $M_0$ for the vd100 model (black solid line) and for its variations with different cross section normalizations that result in a diverse MW satellite population (dark violet), i.e., for the corresponding models shown in dark violet in Fig.~\ref{fig:sigma-M}.
For a model with the normalization of vd100, most of the massive (sub)halos ($>10^9$~M$_\odot$) that are expected to host the MW satellites are predicted to be cored, with only $\sim30\%$ having undergone core-collapse. However, the breadth of the potential collapsed fraction values indicates that even a small re-normalization of the model will significantly increase this fraction. We note that the prediction in Fig.~\ref{fig:fcoll} needs to be tested %
with MW-size simulations with a wide range of cross sections and with enough massive subhalos to sample the high-mass end of the subhalo population, and subsequently recalibrated to match the full range of vd100-style models.
 
\begin{figure}
    \includegraphics[width=\linewidth, clip=True]{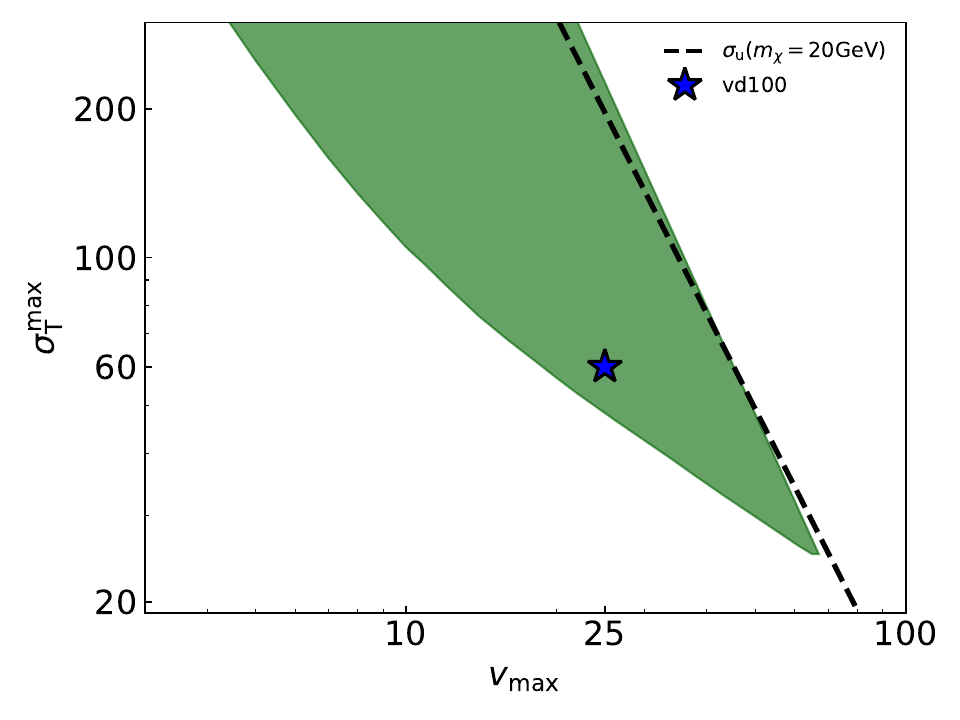}
\caption{The viable values of the parameter space that define the vdSIDM models studied here (Eq.~\ref{eq:cross}): the relative velocity $v_{\rm max}$ where the transfer cross section $\sigma_{\rm T}^{\rm max}$ peaks. 
The highlighted region is that of models that produce a bimodal distribution of halos (green area) and satisfies the constraint from elliptical galaxies $\left<\sigma_{\rm max}\right> \leq 1 \rm cm^2g^{-1}$ for a halo mass of $M_0 = 10^{13}\rm{M_\odot}$. The black dashed line is the unitarity bound for the SIDM cross section for a particle mass $m_\chi = 20 \rm GeV$. The blue star marks the 
vd100 model.}
\label{fig:sigma-vmax}
\end{figure}
In principle, a complete exploration of the Yukawa-like interaction SIDM model in the classical regime requires variations of two parameters,
$v_{\rm max}$ and the normalization $\sigma_{\rm T}^{\rm max}$. 
In addition, the family of SIDM models that generate gravothermally collapsed halos as well as cored halos is restricted by a couple of additional factors we have yet to consider. First, the impact of SIDM in DM structures at larger scales can be compared with observational probes of morphology based on lensing and X-rays. These observations have constrained the transfer cross section
$\sigma_{\rm T}/m_\chi\lesssim0.1-1$~cm$^2$g$^{-1}$ on massive galaxies and galaxy cluster scales (see Section~\ref{sec_intro} for references). In particular, we take the approximate constraint on the cross section in massive ellipticals set in \cite{Peter2013} and more recently in \cite{Despali2022} of $\left<\sigma_{\rm max}\right> \leq 1 \rm cm^2g^{-1}$ at halo mass $M_0 = 10^{13}\rm{M_\odot}$. Second, the SIDM cross section is subject to an upper limit set by the quantum zero-energy resonance, known as the unitarity bound \cite[e.g.][]{Kamada2020}. When the cross section saturates the unitarity bound, it is parameterized solely by the DM mass and is given by:
\begin{equation}
\sigma_{\rm u} = \dfrac{4 \pi}{k^2}
\label{eq:sigma-unitarity}
\end{equation}
where the relative momentum of the scattering particles $k$ is defined as $k=(m/2) v$. We can then use $\sigma_{\rm u}$ to set an upper bound for the cross section $\sigma_{\rm T}^{\rm max}$ that peaks at $v_{\rm max}$:
\begin{equation}
(\sigma_{\rm T} v)\leq \sigma_{\rm T}^{\rm max} v_{\rm max} \leq \sigma_{\rm u} v_{\rm max}
\label{eq:unitarity-bound}
\end{equation}

In Fig.~\ref{fig:sigma-vmax} we show the range of viable parameter space, in peak relative velocity $v_{\rm max}$ and normalization $\sigma_{\rm T}^{\rm max}$, of the Yukawa-like SIDM models (Eq.~\ref{eq:cross}) that satisfy two conditions: a diverse halo population (i.e. cored and gravothermally-collapsed halos) in the range of masses suitable to host the MW satellite population, and that satisfy the constraint from elliptical galaxies. The unitarity bound (for $m_\chi=20$~GeV as a reference) is marked as a dashed line and the pair of $\sigma_{\rm T}^{\rm max}$ and $v_{\rm max}$ corresponding to vd100 is marked with the blue star.

\section{Conclusions} \label{sec_conclusions}

If dark matter (DM) is made of particles that can strongly self-interact and therefore can be described as self-interacting DM (SIDM), the non-linear evolution of DM halos consists of two phases: a cusp-core transformation in which the originally cuspy halo develops a central isothermal core that is in quasi-equilibrium and remains in cored configuration for several \rm{Gyr} 
followed by a rapid gravothermal core-collapse phase. The ultimate consequence of this SIDM-driven collapse is the formation of a black hole with a mass that is a fraction of the central DM core mass \cite{Balberg01}. If the cross section is strongly velocity dependent -- and thus halo mass dependent -- then the population of halos today is expected to exhibit a wide diversity of central density profiles. This behaviour has been invoked as a viable way to explain the diversity of inner DM densities in the Milky Way (MW) satellites \cite{Zavala2019,Correa2021,Correa2022}.

Velocity-dependent SIDM models are the natural result of several particle physics models, e.g. those with new light mediators that produce an effective Yukawa-like interaction between DM particles (for a review see e.g. \cite{Tulin2018}). These models also have the advantage of avoiding the stringent constraints on the cross section from gravitational lensing, X-ray morphology and dynamical analysis in cluster mergers and elliptical galaxies, which limits $\sigma_{\rm T}/m_\chi\lesssim1$~cm$^2$g$^{-1}$ at these scales \cite{Peter2013,Robertson2017,Robertson2019,Harvey2019,Andrade2022,Eckert2022,Shen2022,Despali2022}, well below the threshold for gravothermal collapse. In this SIDM scenario, the structure of DM today is indistinguishable from the cold dark matter (CDM) scenario at cluster scales. The deviation from CDM grows at smaller scales, starting with the development of spherical isothermal cores in the center of $10^{11}-10^{12}$\msun halos, and followed by the onset of gravothermal collapse for the halos of dwarf spheroidal galaxies, a fraction of which should host SIDM-generated intermediate mass black holes (IMBHs).

In this work, we develop an analytical framework to predict the timescales and mass scales for the formation of IMBHs in SIDM halos, which includes the different stages in the cusp-core-collapse evolution of SIDM halos in a cosmological setting (Section~\ref{sec_SIDM-halos}). This framework is calibrated to a high-resolution simulation of a benchmark velocity-dependent SIDM model (vd100; see Fig.~\ref{fig:cross_section}), which has been tuned to produce a large diversity in the MW satellite population \cite{Zavala2019, Turner, Lovell2022}. Our main results are summarized as follows:

\begin{itemize}

\item The black hole formation (gravothermal collapse) timescale as a function of present day halo mass (in isolation) is shown in Fig.~\ref{fig:formation_time}. We consider two starting redshifts that bracket the range of the possible threshold epochs ($z_{\rm cc}$) when the cusp-core-collapse evolution starts (see Section~\ref{sec_redshift}). The 
assembly/formation redshift of the primordial CDM NFW halo $z_{\rm cc}=z_{\rm form}=z_{-2}$ (yellow lines; defined as the time of assembly of the central region of the halo within its scale radius according to the model by Ref.~\cite{Ludlow2016}), and the redshift of the last major merger of the halo $z_{\rm cc}=z_{\rm LMM}$ (blue lines; takes into account cosmological accretion and it is computed using the simulation results of \cite{Fakhouri}). We also consider the possible acceleration of the collapse timescale driven by tidal stripping following \cite{Nishikawa} (dashed lines in Fig.~\ref{fig:formation_time}; see Section~\ref{sec_tidal-stripping}).

\item We compare our results with the mass threshold ($\sim3\times10^8 \rm M_\odot$), where most of the MW (sub)halos are observed to have undergone gravothermal collapse according to the vd100 SIDM simulation analyses made in \cite{Turner} and more recently in \cite{Lovell2022}. We find that our modelling can be fitted to this result by a small re-calibration of the free parameter $C$ in the gravothermal fluid formalism (Eq.\ref{eq:t_rel}) by choosing $z_{\rm cc}=z_{\rm LMM}$ and by assuming that tidal stripping has no impact on the collapse time (see Fig.~\ref{fig:viable-mass-range}). This choice is seemingly consistent with previous expectations that the core phase is delayed by cosmological infall \cite{Ahn2005}, and supports the recent detailed simulation work by \cite{Zeng2022} that suggests the impact of tidal stripping in accelerating core collapse is likely overestimated in \cite{Nishikawa} -- from which we developed our incorporate tidal stripping model -- due to the competing environmental effect of subhalo heating through collisions with host halo particles.

\item We show the black hole mass $M_{\rm{BH}}$ as a function of the present day halo mass $M_0$ in Fig.~\ref{fig:M0-Mbh} (violet and red lines). This estimated seed black hole mass is obtained with Eq.~\ref{eq:M_bh}, which is derived by following the evolution of the part of the core that collapses to high central densities within the gravothermal fluid formalism. The development of the relativistic instability ultimately leads to the formation of the black hole \cite{Balberg02}. Remarkably, the slope of the $M_{\rm BH}-M_0$ SIDM-core-collapse relation is similar to that of the extrapolated SMBH--halo mass empirical relation found in massive galaxies \cite{Marasco2021}. This mechanism could then potentially constitute a continuation of the empirical relation towards the regime of dSphs, although the normalization is two orders of magnitude below the expectation, and thus the seed SIDM-driven black hole would need to grow substantially to satisfy such a scenario and would have to be rapidly accelerated compared to our predictions.

\item We also consider the impact of the cosmological scatter in the concentration--mass relation in our results by first considering a 10\% scatter on the concentration at a fixed mass (today) \cite[e.g.][]{SC2014}. We find that more concentrated halos have their core collapse several {\rm Gyrs} earlier (shaded areas in Fig.~\ref{fig:formation_time}). The fraction of halos that is expected to collapse strongly depends on the range of concentrations available to the halo population at a given mass (see Fig.~\ref{fig:fcoll}). The predicted black hole mass is however, mainly set by halo mass, and is only weakly affected by concentration (at the percent level).

\item Finally, we estimated the range of self-scattering cross sections that result in a diverse MW satellite population, i.e., that generate both cored and core-collapsed host halos for MW satellites. We first consider SIDM models with the same velocity dependence as vd100 but with different normalization (see Eq.~\ref{eq:cross}). The results are shown in Figs.~\ref{fig:sigma-M} and \ref{fig:fcoll}. We found that the vd100 model has a normalization that is close to the lower limit to exhibit diversity with most of the subhalos in MW-size systems being cored, especially for halos $M_{0}>10^{10}$\msun, of which $\sim70$\% are cored. The latter is expected since our default choice of parameters is calibrated to the recent results in \cite{Lovell2022}, who agree qualitatively on the scarcity of massive core-collapsed subhalos in vd100. As noted by \cite{Lovell2022}, this represents a potential issue with models such as vd100 since it would be more natural to expect the bright MW dSphs to be assigned to the most massive subhalos, which in this case are likely to be cored, and thus inconsistent with the properties of dense bright dSphs. We show that such a potential issue can be alleviated by a relatively small increase in the cross section normalization of the vd100 model, since the fraction of massive core-collapsed halos increases rapidly with larger cross sections (see Fig.~\ref{fig:fcoll}). To expand upon this result, we explore in Fig.~\ref{fig:sigma-vmax} the parameter space of the Yukawa-like SIDM model (Eq.~\ref{eq:cross}) to find the viable values of the peak velocity $v_{\rm max}$ and the normalization $\sigma_{\rm T}^{\rm max}$ that satisfy: i) having a diverse MW satellite population (as described before); ii) satisfy constraints at larger scales, in particular those set by elliptical galaxies \cite{Peter2013,Despali2022}; and iii) satisfy the constraint set on the cross section by the unitarity bound. These conditions set an upper bound for $v_{\rm max}$ to be $v_{\rm max}<70 \rm kms^{-1}$ with a very narrow range of possible $\sigma_{\rm T}^{\rm max}$ values, while an ever wider range of $\sigma_{\rm T}^{\rm max}$ is suitable for smaller $v_{\rm max}$ values down to $v_{\rm max}\sim 5 \rm kms^{-1}$. To ascertain whether a single SIDM model can fit all MW satellites simultaneously requires much more precise estimates for the fraction of collapsed-to-total MW satellites.
\end{itemize}

In Appendix \ref{calibration}, we discuss different values of calibration parameter $C$ in the gravothermal fluid model and their impact on our results. Based on previous studies this parameter has a plausible range between $0.4-0.75$, with our calibrated value being at the low end. However, such a value seems to be appropriate/favored by analyses based on cosmological simulation \cite{Yang2022} (see also \cite{Essig2019}). Nevertheless, we explore in Appendix \ref{calibration} how choosing a value at the high end $C=0.75$ would impact our key results. We find that although the timescale for black hole formation significantly changes with such a value of $C$, (see Fig.~\ref{fig:time075}), the impact on our key result is modest, that is the region in the plane $\sigma_{\rm max}-M_0$ that would result in a diverse (i.e. cuspy and cored) MW satellite population, is shifted downward overall by about a factor of $\sim1.5$ in the normalization of the cross section (see Fig.~\ref{fig:Onset075}).

If SIDM is to be invoked to explain the diversity of the dSph MW satellite population, then the presence of IMBHs in the center of cored-collapsed (cuspy) halos is unavoidable. Based on our work, the range of seed IMBH masses for the dSph MW satellites that can potentially be hosted by cored-collapsed SIDM halos is in the range $0.1-1000$~M$_\odot$ (Fig.~\ref{fig:sigma-M}).
The existence of such IMBHs may be verified by detailed observations of dSph kinematics with the upcoming generations of extremely large telescopes \cite{Greene2020}. Due to their high spatial resolution, they should be able to find  $<10^5$\msun  black holes in $<10^9$\msun 
in nearby galaxies through high-precision proper motion measurements \cite{Greene2020}. For example, to reach the range of $10^3-10^4$\msun Ref.~\cite{MacLeod2016} suggest looking at the disruption of stars passing close to IMBHs. Under certain conditions, IMBHs should typically acquire companions with orbital periods of years, corresponding to semi-major axes of $\sim 5-10$ mas for $\sim 10^3 M_\odot$ IMBHs.

More precise observational kinematic data for the dSphs in the future will constraint further the inner DM content of these systems, possibly establishing their inner DM profile. This will conclusively determine the significance of the diversity problem and constrain different classes of solutions. On the theoretical side, the predictions of vdSIDM models such as those considered in this work need to be complemented with more dedicated simulations that explore the relevant parameter space of cross sections that contain the gravothermal collapse regime \cite{Zavala2019,Correa2022}, and can be complemented with a semi-analytical revision of the predicted seed black hole mass in the seminal work by Ref.~\cite{Balberg01} within a full cosmological setting. Finally, a key aspect to explore is the interplay between the formation and evolution of the visible baryonic galaxy with the collapsing SIDM core and its central black hole.

\begin{acknowledgments}
TM, JZ, and MRL acknowledge support by a Project Grant from the Icelandic Research Fund (grant number 206930).
    
\end{acknowledgments}

\newpage

\appendix

\bibliography{Gravothermal_Collapse}

\appendix
\section{}\label{sec_vel_independent_t_r}

\begin{figure}
    \includegraphics[width=\linewidth]{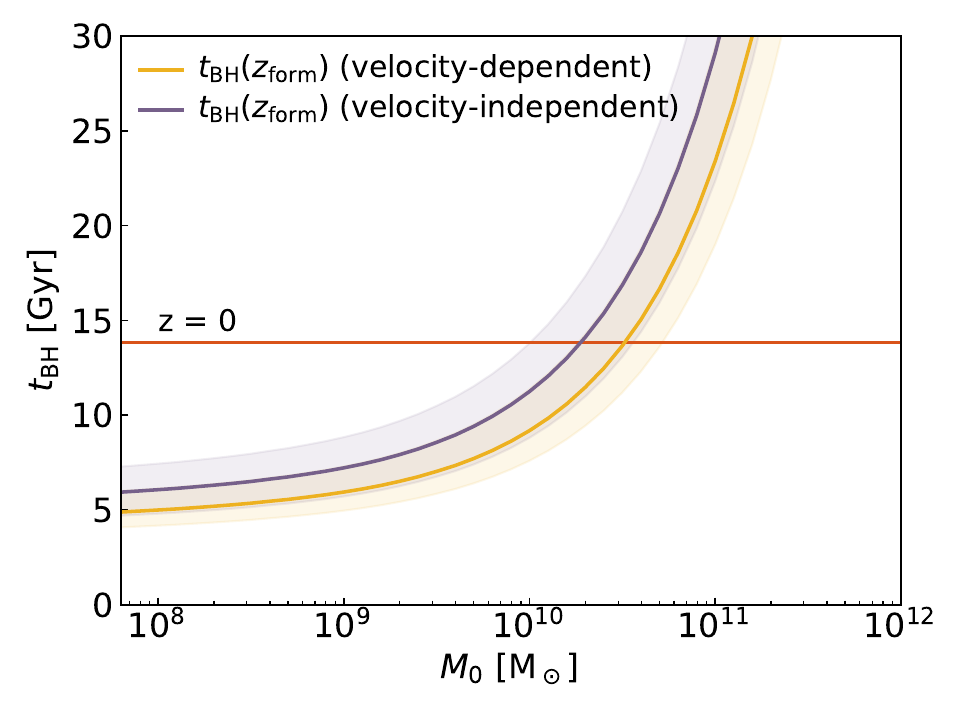}
\caption{Timescale to form a black hole in SIDM halos as a function of the present-day halo mass $M_0$ (in isolation). The yellow line (Eq.~\ref{eq:collapse_time}) represents the cusp-core-collapse evolution timescale chosen for this work (equal to the curve of the same color in Fig.~\ref{fig:formation_time}), and the violet line (Eq.~\ref{eq:collapse_time_pollack}) is based on \cite{Pollack}.
As in Fig.~\ref{fig:formation_time}, shaded bands indicate a concentration scatter of $\pm10\%$ in the concentration-mass relation \cite[e.g.][]{SC2014}. The horizontal line indicates the age of the Universe at $z=0$.}
\label{fig:vel-independent}
\end{figure}

In this work, we computed the relaxation timescales for self-scattering and core-collapse using Eqs.~\ref{eq:t_rel} and ~\ref{t_col} respectively. These equations are adapted from \cite{Outmezguine2022}, which incorporates a velocity-dependent SIDM cross section into the gravothermal fluid formalism. In this Appendix, we consider the impact on our results of using an alternative set of formulae developed by Ref.~\cite{Pollack}, where the SIDM cross section is assumed to be constant. According to this reference, the cusp-core-collapse timescale is given by:
\begin{equation}
t_{\rm coll}=455.65t_r(z_{\rm cc}),
\label{eq:cc_collapse}
\end{equation} 
where $t_r(z_{\rm cc})$ is the relaxation time derived for velocity-independent cross sections:

\begin{equation}
\begin{split}
t_r(&z_{\rm cc}) = \dfrac{1}{a\sigma_{\rm max}}\left(\dfrac{k(c)^2}{4\pi \rm G^3}\right)^{1/6}\delta_{\rm char}^{-7/6}\rho_{\rm crit}(z_{\rm cc})^{-7/6}M_{0}^{-1/3} \\[5pt]
&= \mathrm{0.310\ Myr} \times\left(\dfrac{M_{0}}{10^{12}\rm{M_{\odot}}}\right)^{-1/3} \left(\dfrac{k(c(z_{\rm cc}))}{k(9)}\right)^{3/2} \\[5pt]
&\left(\dfrac{c(z_{\rm cc})}{9}\right)^{-7/2} \left(\dfrac{\rho_{\rm crit}(z_{\rm cc})}{\rho_{crit}(z=15)}\right)^{-7/6}\left(\dfrac{\left<\sigma_{\rm max}\right>}{1\mathrm{\ cm^2/g}}\right)^{-1}.
\end{split}
\label{eq:t_r_pollack}
\end{equation}

Then the time to form a black hole is given by:
\begin{equation}
t_{\rm BH}(M_0,\sigma_{\rm max})=t(z_{\rm cc})+455.65t_r(z_{\rm cc}).
\label{eq:collapse_time_pollack}
\end{equation} 

In Fig.~\ref{fig:vel-independent} we compare the black hole formation timescale derived self-consistently for velocity-dependent cross sections (Eq.~\ref{eq:collapse_time}) with the case derived assuming the velocity-independent cross section (Eq.~\ref{eq:collapse_time_pollack}). For this comparison, we have used the case with $z_{\rm cc}=z_{\rm form}=z_{-2}$ un-calibrated and without tidal stripping (i.e. equivalent to the solid yellow line in Fig.~\ref{fig:formation_time}). The $t_{\rm BH}-M_0$ curves have a similar shape in both cases across all the explored mass range, but with a $\sim1$~Gyr difference in normalization. For the purposes of this work, such a difference could be absorbed almost completely in the calibration factor $C$ in Eq.~\ref{t_col}. 

\section{}\label{calibration}
In the gravothermal fluid formalism, the heat conductivity $\kappa$ 
is a key quantity in the time evolution of the SIDM halo. In the LMFP regime, it can be derived by dimensional analysis but it carries an
unknown parameter $C$ that is the order of unity, which cannot be derived from first principles \cite[e.g.][]{Balberg01, Koda2011, Pollack, Essig2019, Nishikawa, Outmezguine2022}. To determine $C$ all these studies compared the evolution of the halo density profile given by the gravothermal fluid model to that obtained from different types of $N-$body simulations, where hard-sphere elastic scattering interactions were implemented. Most of these studies used isolated idealized simulations with only Refs.~\cite{Essig2019,Nishikawa, Outmezguine2022} using the cosmological constant-cross-section SIDM simulations presented in \cite{Elbert2015} and \cite{Koda2011}. For large cross sections that have entered the regime of core-collapse, the latter studies suggest that a value of $C=0.45$ is a better fit to cosmological simulations, while previous analyses based on isolated simulations preferred values around $0.6-0.75$.

In this work, we used the formula for the relaxation time (Eq.~\ref{eq:t_rel}) and, initially, the fitting parameter $C=0.57$ calibrated for the velocity-dependent SIDM model in Ref.~\cite{Outmezguine2022} using isolated simulations. This value was roughly in between the values explored in the literature as discussed above. After comparing the timescale for gravothermal collapse (black hole formation) with the results from \cite{Lovell2022} based on the high-resolution velocity-dependent SIDM simulation performed in \cite{Zavala2019}, we found that a value of $C=0.42$ is a more accurate fit (see Fig.~\ref{fig:viable-mass-range}). This is in qualitative agreement with the analysis in \cite{Essig2019}, who found $C=0.45$ consistent with the cosmological simulation they analyzed. However, we note that there are different methods used for calibration. Our calibration is based on comparing the mass threshold where $\geq 50\%$ subhalos of a MW-size host have collapsed, while \cite{Essig2019} uses the density profile of dwarf-size main host halo. More recently, a new velocity-dependent SIDM cosmological simulation with a similar resolution to the one we used has been performed \cite{Yang2022}. Using a similar comparison to ours, albeit with a different methodology, between the gravothermal fluid model and the simulation, they find that a value of $C=0.75$ overestimates the number of subhalos that should collapse by $z=0$ by around a factor of 2. Thus, this result also suggests that a lower value of $C$ fits cosmological simulations better, which is qualitatively in agreement with our finding.

\begin{figure}
    \includegraphics[width=\linewidth, trim=0cm 0cm 0.2cm 0cm]{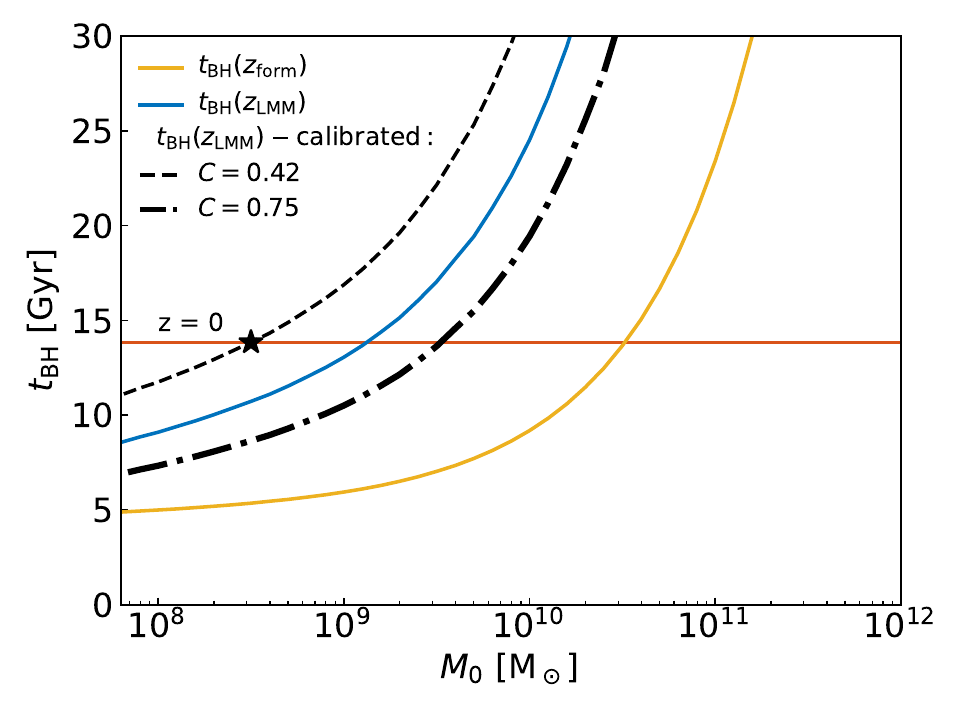}
\caption{The same as Fig.~\ref{fig:viable-mass-range}, but with the addition of the thick dot-dashed black line, which is the same as the calibrated model used for our key results (thin dashed black line) but with $C=0.75$ instead.}
\label{fig:time075}
\end{figure}
\begin{figure}
    \includegraphics[width=\linewidth, trim=0cm 0cm 1.2cm -0.6cm]{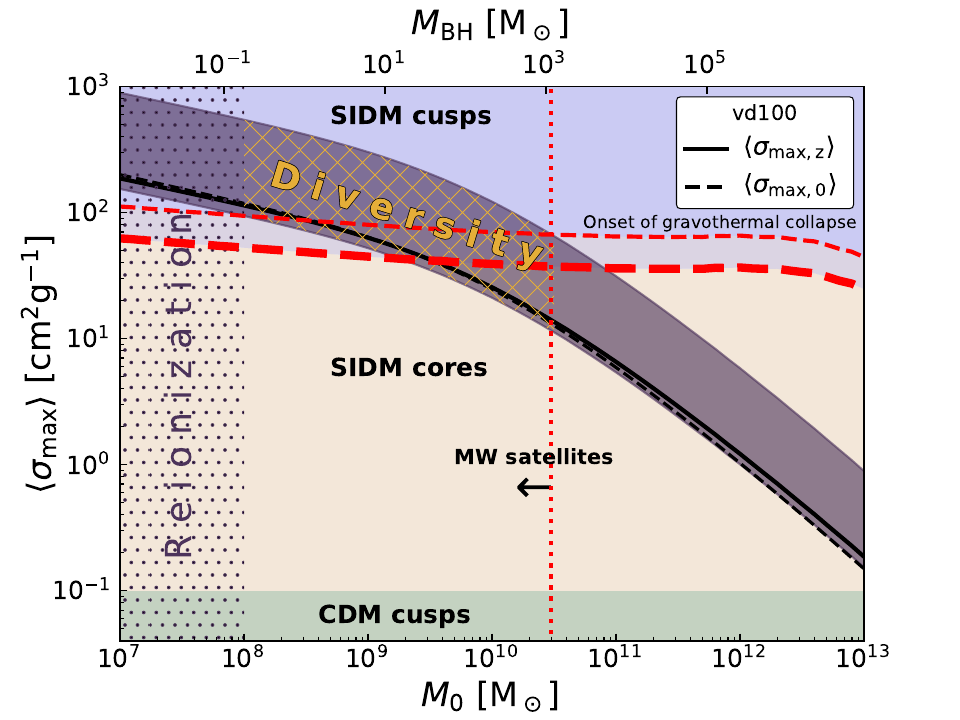}
\caption{The same as Fig.~\ref{fig:sigma-M}, but with the addition of the nearly horizontal thick red dashed line -- the required cross section value for the onset of gravothermal collapse when $C=0.75$ is used instead of $C=0.42$, which is the calibrated value used for our key results (thin red dashed line).}
\label{fig:Onset075}
\end{figure}

In order to obtain a precise value of the parameter $C$ that is appropriate for subhalos in a cosmological setting, it would be necessary to perform a detailed analysis across multiple high-resolution simulations that considers variations across different velocity-dependent cross sections and a large exploration across different host halo masses. This hypothetical study requires many more simulations than have been performed to date; currently only two exhibit the required mass resolution (\cite{Zavala2019} and \cite{Yang2022}). Such an analysis is beyond the scope of this paper. Nevertheless, we can take into account the uncertainty in the value of $C$ by considering its impact on the key results of our work. We do this in this appendix by comparing the predictions of our model with the calibrated $C=0.42$ value we have used in our key results, and with a value of $C=0.75$. The latter is the highest value for this parameter explored in \cite{Balberg01,Nishikawa} and as we argued before is at the high end of plausible values as found in \cite{Yang2022}.

\begin{figure}
    \includegraphics[width=\linewidth, trim=0cm 0cm 0.2cm 0cm]{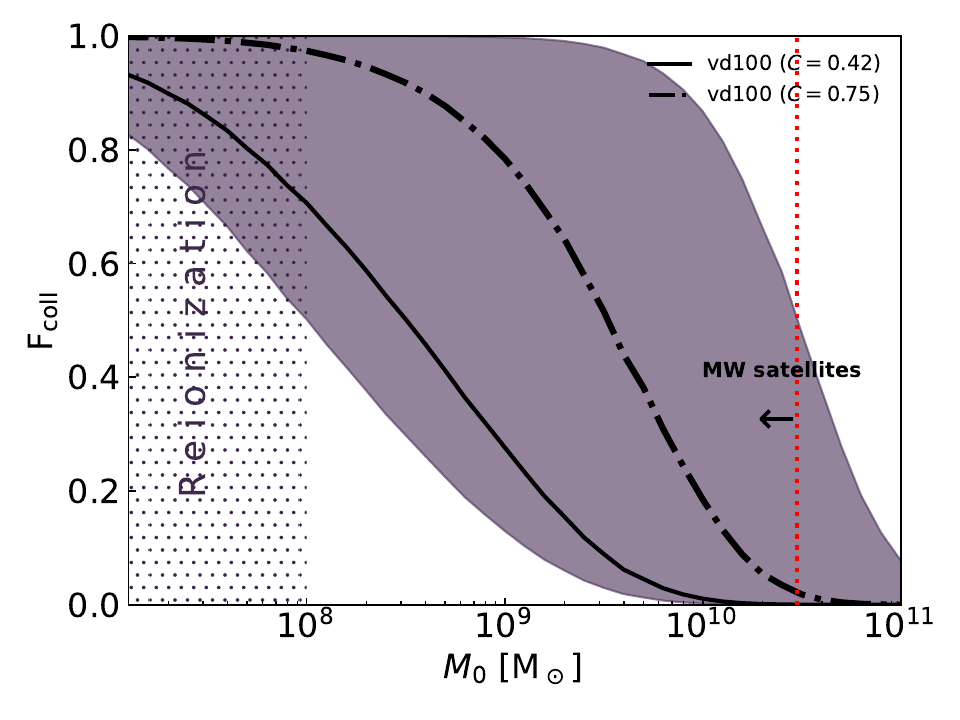}
\caption{The same as Fig.~\ref{fig:fcoll}, but with the addition of the thick black dot-dashed line -- the fraction of core-collapsed halos as a function of $M_0$ when $C=0.75$ is used instead of $C=42$, as in our key result (solid black line).}
\label{fig:fcoll075}
\end{figure}

The impact on the timescale for black hole formation due to variations in $C$ across this plausible range of values is shown in Fig.~\ref{fig:time075}, specifically as the range between the dashed and thick dot-dashed black lines. We note that the resulting range is similar to the spread observed due to the $10\%$ cosmological scatter in the concentration--mass relation (see Fig.~\ref{fig:formation_time}) with an increase in the threshold mass that divides subhalos that have collapsed from those that have not by about an order of magnitude. 
Naturally, setting a higher value for the $C$ parameter lowers the cross section values required to reach the threshold for the onset of gravothermal collapse by about a factor of 1.5; this is shown in Fig.~\ref{fig:Onset075} through the difference between the thin and thick red dashed lines. The overall change would then be a shift downwards of the region that is expected to have a diverse MW subhalo satellite population of cores and cusps. We conclude that the overall impact of the $C$ parameter uncertainty on our key plot Fig.~\ref{fig:sigma-M} is a factor of $\sim1.5$ on the normalization of the transfer cross section that results in a diverse MW satellite population.
Finally, we show in Fig.~\ref{fig:fcoll075} the fraction of core-collapsed halos as a function of $M_0$ when setting $C=0.75$ instead of our default value of $C=0.42$ as a thick dot-dashed black line. With this highest value of $C$, $80\%$ of the (sub)halos with $>10^9 M_\odot$ are expected to have undergone core collapse. This is around $50\%$ higher than in the case of $C=0.42$. Setting other values of $C$ in the plausible range would result in different core-collapsed halo fractions in between the two black solid and dot-dashed lines.

\end{document}